%% file: main.tex
\documentstyle[12pt,psfig]{article}
\input{sty}

\title{
Anomalous size--dependence of interfacial\\ 
profiles between coexisting phases\\
of polymer mixtures in thin film geometry:\\
A Monte--Carlo simulation
}
\author{
{\bf Andreas Werner, Friederike Schmid, Marcus M\"uller,}\\
{\bf and Kurt Binder}\\[5mm]
{\it Institut f\"ur Physik, Johannes Gutenberg Universit\"at Mainz}\\
{\it Staudinger Weg 7, D-55099 Mainz, Germany}
}
\date{
\today\\[2cm]
to be appear in JCP, 15 Nov 1997
}
\begin{document}
\include{text}
\include{figures}

\include{bib}
\end{document}

%% file: sty.tex
\newcommand{\bi}{\begin{itemize}}
\newcommand{\ei}{\end{itemize}}
\newcommand{\bd}{\begin{description}}
\newcommand{\ed}{\end{description}}
\newcommand{\ra}{$\rightarrow$\hspace{3mm}}

\newcommand{\erf}{\mbox{erf}}
\newcommand{\vect}[1] { \mbox {\boldmath $ #1 $} }

\newcommand{\kt}{k\indice{B}T}
\newcommand{\beq}{\begin{equation}}
\newcommand{\eeq}{\end{equation}}
\newcommand{\bea}{\begin{eqnarray}}
\newcommand{\beas}{\begin{eqnarray*}}
\newcommand{\eea}{\end{eqnarray}}
\newcommand{\eeas}{\end{eqnarray*}}

\newcommand{\indice}[1]{_{\rm \scriptscriptstyle #1}}

%% file: text.tex
\maketitle
\newpage
\begin{abstract}
The interfacial profile between coexisting phases of a binary mixture (A,B)
in a thin film of thickness $D$ and lateral linear dimensions $L$
depends sensitively on both linear dimensions and on the nature of
boundary conditions and statistical ensembles applied.
These phenomena generic for systems in confined geometry
are demonstrated by Monte--Carlo simulations of the bond fluctuation model
of symmetric polymer mixtures, using chains containing
$N_A = N_B = N = 32$ effective monomers connected by effective bonds
with an attractive interaction between monomers of the same type and
a repulsive interaction between different types.
We use short range potentials at the walls, the right wall favoring A monomers
and the left wall B monomers.
Periodic boundary conditions are applied in the directions parallel to the 
walls.
Both the canonical and semi-grand-canonical ensemble are studied.
We argue that the latter case is appropriate for experiments with a lateral
resolution $L$ much less than the actual lateral sample size, in thermal
equilibrium.
In the canonical ensemble, the interfacial width $w$ increases
(from small values which are of the same order as the ``intrinsic profile'')
like $w \propto \sqrt{D}$, before a crossover to a saturation value
$w_{max}$ ($w_{max}^2 \propto \ln L$) sets in.
In the semi-grand-canonical ensemble, however, one finds the same widths
$w \propto \sqrt{D}$ as in the canonical ensemble for not too large $L$,
while for large $L$ the interfacial profile is smeared out over a finite
fraction of the film thickness ($w \propto D$ for $D \to \infty$). 
We discuss the implications
of these findings for the interpretation of both simulations and experiments.
\end{abstract}
\newpage
\section{Introduction}
\label{s1}
Since van der Waals' famous treatment of the interface between coexisting
fluid and gas, calculations and measurements of interfacial profiles
between coexisting phases have remained a particular 
challenge~\cite{1,2,3,4,5}.
Van der Waals type mean field theories were extended to 
mixtures by Cahn and Hilliard~\cite{6}, as well as to more complex systems 
such as polymer solutions~\cite{7}, polymer blends~\cite{8,9,9b}, 
microemulsions~\cite{10,11}, etc.
These theories, but also more complex ones such as
the self-consistent field approach to strongly segregated polymer 
mixtures~\cite{12,13,14,15,16,17,18,19,20,21} consider a
hypothetical ``intrinsic'' profile that is neither directly accessible
by experiment nor by computer simulation. 
I.e., both in experiment and in simulations these ``intrinsic'' profiles ---
if they have a well--defined meaning at all --- are broadened by fluctuations 
(such as capillary waves~\cite{18,19,22} which can be modelled by harmonic 
displacements of the local interface position~\cite{1,2,3,4}), 
but this broadening is also limited by the geometry of the (finite~!) system 
that is considered and affected by forces due to the boundaries of the system.
In fact, the latter may give rise to interface binding/unbinding phenomena
(``wetting''~\cite{23,24,25,26,27,28,29}),
and the associated fluctuations do have a strong effect on the observable
average interfacial profiles~\cite{30,31,32,33,33b}.
(Note that external potentials such as gravity, which clearly limit
interfacial fluctuations in liquid--gas systems, are disregarded here
throughout.)

There are numerous examples in the literature, both in experiment (e.g.
on polymer mixtures~\cite{34,35}) and simulation (ranging from lattice gas
models~\cite{36} to the water--vapor interface~\cite{37}) where
this fluctuation broadening of interfacial profiles is disregarded and
the resulting interfacial width is interpreted as if the intrinsic width
were obtained.
Recent work~\cite{33}, however, indicates that such an approach may yield
rather misleading results~!

Fig.1 sketches the $L \times L \times D$ geometries 
that are typically used in simulations~\cite{22,31,32,33,33b,36,37} and
experiments~\cite{33,34,35}.
$L$ is the linear dimension parallel to the interface(s), 
$D$ perpendicular to the interface(s).
Fig.1a shows a system with a single interface between two demixed phases
using either hard walls as boundary condition normal to the
interface (left part) or ``antiperiodic'' boundary conditions
(i.e., an A monomer leaving the right boundary of the simulation box
reenters at the left boundary as a B monomer, and vice versa; right part).
This antiperiodic boundary condition is particular useful for strictly
symmetrical mixtures where two species A,B intrinsically are not different
apart from their ``label'', and thus one can achieve a geometry with a
single interface and no disturbing effects due to external walls.
If periodic boundary conditions are used in all three directions,
the system must contain two interfaces (Fig.1b).  
We emphasize that the concentration profile (and hence also its width $w$)
obtained in these situations depends both on $D$ and on $L$, and in general
may differ appreciably from any sensibly defined intrinsic profile~!
The present paper intends to study these dependences for a simple model
system that is nevertheless close to physical systems of practical
interest, namely polymer mixtures~\cite{18,33,34,35}.
Experiments are done in the geometry of Fig.1c and encounter closely
related problems of interpretation as the simulations as will be explained
below.

While in the thermodynamic limit the different ensembles of statistical
mechanics yield equivalent results, this is not true in systems which have
all their linear dimensions finite~\cite{38}.
Considering binary (A,B) mixtures at constant density, we may
deal either with the canonical ensemble where the average volume fraction
$\overline{\Phi}$ of species A is held fixed 
(if $\overline{\rho}_A,\overline{\rho}_B$
are the average densities of A~particles/B~particles in the system,
then $\overline{\Phi}_A = \overline{\Phi} = \overline{\rho}_A 
/ (\overline{\rho}_A + \overline{\rho}_B)$, and
$\overline{\Phi}_B = 1 - \overline{\Phi} = \overline{\rho}_B 
/ (\overline{\rho}_A + \overline{\rho}_B)$
being the volume fraction of species B)
or with the semi-grand-canonical ensemble where the chemical potential
difference $\Delta \mu$ between A and B is held fixed.
Note that in the left part of Fig.1a the use of the canonical ensemble
suppresses the fluctuation of the average position of the interface ---
e.g. for $\overline{\Phi} = 1/2$ and ``antisymmetric'' walls acting 
on a ``symmetric'' mixture~\cite{39} (as considered here and in
Refs.~\cite{31,32,36}) the average position of the interface is fixed
in the middle of the film, $z = D/2$.
(We disregard here the case~\cite{30,31,32} of temperatures below the 
transition temperature $T_c (D)$ of the interface 
localization--delocalization
transition where the interface is bound either to the left or the right
wall and $\overline{\Phi} = 1/2$ is maintained only via a lateral
inhomogeneity, i.e., formation of domains
separated by interfaces running perpendicular to the walls~\cite{40}.)
In contrast, using the semi-grand-canonical ensemble, $\overline{\Phi}$
and hence the average interface positions are fluctuating
and, in particular for small $L$, this fluctuation of the average interface
position may even be more important than local distortions of the interface.
Naturally, one has to be careful in the interpretation of experiments, too:
While the canonical ensemble is appropriate for a discussion on the length
scale $L_{sample}$ comprising the total sample, this is not true if the
measurement is done with a technique characterized by
a length scale $L$ of lateral resolution with $L \ll L_{sample}$
(Fig.1c).
The condition $\overline{\Phi} = const$ fixes the average interface
position on the length scale $L_{sample}$, but it clearly does
not fix it on the length scale $L$ that is probed.
This is the case for experiments where interfacial properties 
are measured ``in situ'' in the melt phase~\cite{40b}.
We may consider the slab of width $L$ that is measured as a ``subsystem''
in a statistical sense that can exchange A,B particles freely by 
diffusion with the rest of the system. This acts as a reservoir,
in thermal equilibrium, and hence the semi-grand-canonical ensemble applies.
Implicit in this interpretation is the assumption that the time
constant of the measurement is large enough such that one
measures not just a kind of ``snapshot'' of the interfacial configuration,
but carries out a meaningful statistical average in thermal equilibrium.

For the fully periodic system~\cite{33b,37} (Fig.1b), 
on the other hand, semi-grand-canonical
techniques are inappropriate since then the two-phase configuration shown
is at best meta-stable: If one simulated long enough, the two interfaces
would meet and annihilate each other, and either the A-rich domain or the
B-rich domain would vanish altogether.
The condition $\overline{\Phi} = const$ in the canonical ensemble
and fixing the center of mass of the B-rich phase then suppresses also 
fluctuations of the average interface positions.
Finally, in the antiperiodic case there is a clear translational invariance;
no origin at the z-axis is distinguished except by the location of the
interface itself.
Then it is natural to record all information in a moving frame:
In each configuration that is analyzed the point $z=0$ is chosen to
coincide with the interface position averaged in x,y directions~\cite{22,41}.
Even in the semi-grand-canonical ensemble, the fluctuation of the average 
interface position is thus eliminated per construction.

In the present paper we shall restrict our attention
to the case of a single
interface confined between ``antisymmetric'' walls (i.e., the left part
of Fig.1a)
since this situation is closest to the experimental one (Fig.1c), 
studying the finite size effects both in the canonical and the 
semi-grand-canonical case.
We anticipate, however, that many of our findings carry over qualitatively
to the other cases (antiperiodic boundary conditions in Fig.1a, 
periodic boundary conditions in Fig.1b) as well.

In the next section, we define the model that is used in our study
and make a few remarks on the simulation technique.
Sec.\ref{s3} briefly reviews the theoretical background,
while Sec.\ref{s4} gives our results on interfacial profiles and
interfacial widths and discusses the correlation function of interfacial
fluctuations in directions parallel to the walls and the associated correlation
length $\xi_{||}$.
Sec.\ref{s5} describes a test of capillary wave concepts.
Finally, Sec.\ref{s6} summarizes our conclusions.
\section{Model and simulation technique}
\label{s2}
Since we desire to make as close contact as possible with experiments
on polymer mixtures confined in thin film geometry,
where size effects as discussed here have already been 
observed~\cite{33,42}, we study a coarse--grained model for
a symmetric polymer mixture, extending previous studies of the bond 
fluctuation model~\cite{43} on the simple cubic 
lattice~\cite{21,22,44,45,46}.
We summarize here only very briefly some key features~\cite{47}
of this well studied model:
One integrates $n = 3-5$ chemical bonds along the backbone of
a real polymer chain into one effective bond, whose length is
allowed to fluctuate between the following values:
$2,\sqrt{5},\sqrt{6},3,\sqrt{10}$, all lengths being measured in units
of the lattice spacing.	
This variation in the bond length is not only computationally
advantageous~\cite{44,47,48}, but also physically reasonable as a 
result of the internal degrees of freedom of the combined
$3-5$ original chemical bonds.
Furthermore, the 87 different bond angles  between subsequent
effective bonds approximate properties of continuum models
already rather closely.
These effective bonds are joined at effective monomers,
each blocking all 8 sites of an elementary cube from further occupation.
We work at a volume fraction $\Phi_b = 1/2$ of occupied lattice sites,
which corresponds to a dense melt~\cite{48}, i.e., the density
of effective monomers is $\rho_b = 1/16$ in our units.
For consistency with previous work~\cite{22} where interfaces for the 
antiperiodic boundary conditions (Fig.1a, right part) were studied
we choose here a chain length $N = 32$, which hence
corresponds to a degree of polymerization $N_p = nN \approx 96 -160$.
For this choice, the radius of gyration $R_g$ of the chains is about 
$R_g \approx 7$ lattice spacings.
Defining then the effective
statistical segment length $b$ from the relation valid for Gaussian
chains, $R_g = b \sqrt{N/6}$, we obtain $b \approx 3.05$.
Note that the end-to-end distance is $R \approx 17$,
i.e., within our accuracy the relation $R = b \sqrt{N}$ is also
fulfilled.
Of course, it would be nice to study much longer chains so that
the length scales $b$ and $R_g$ are more distinct from each other,
but this would require an unreasonable large effort in computer
time and thus has not been attempted in the present context.

The Hamiltonian is written in an Ising model type spin 
representation~\cite{39} as
\beq
\frac{{\cal H}}{ \kt } = 
- \epsilon \sum_{d(i,j) \le \sqrt{6}} S_i S_j
+ \epsilon_w \sum_{z_i \le 2} S_i - \epsilon_w \sum_{z_i \ge D-2} S_i,
\label{eq1}
\eeq
where $S_i = +1$ if monomer $i$ is of type A,
$S_i = -1$ if it is of type B,
the interaction parameter $\epsilon$ being defined as 
$\kt \epsilon = \epsilon_{AA} = \epsilon_{BB} = -\epsilon_{AB}$,
if the distance $d(i,j)$ between monomers does not exceed $\sqrt{6}$,
while for more distant neighbors the interaction is zero.
From previous work~\cite{45} it is known that the critical point of
phase separations occurs at $\epsilon_{crit} = 0.0144$, and since
we wish to avoid critical fluctuations, we work here at 
$\epsilon = 0.03$, i.e., far below the critical point
($T/T_{crit} = 0.48$), although experiments are done for~\cite{33}
$T/T_{crit} \approx 0.95$.
In the latter case, however, the correlation length $\xi$ of order
parameter fluctuations is already about twice as large as the radius
of gyration, while in our case it is distinctly smaller,
$\xi \approx 3.6$.
Since the correlation length sets the scale for the concentration 
profile --- at least in its wings~\cite{9} --- one would need to simulate
much larger systems for $T/T_{crit} \approx 0.95$ than for the present
choice, and in fact with the present computer facilities it is not
yet possible to work so close to the critical point.
Our choice corresponds  to a Flory--Huggins 
parameter~\cite{49,50} $\chi = 2 z_{eff} \epsilon \approx 0.159$
since the effective coordination number is~\cite{22} 
$z_{eff} \approx 2.65$, which is still physically reasonable although
it corresponds to a strongly segregated situation ($\chi N \approx 5.09$).
From previous work the interfacial tension $\Sigma$ is known~\cite{22}
as well; namely $\Sigma / \kt = \sigma \approx 0.015$.

In Eq.\ref{eq1}, a wall interaction of square-well type has been added
if monomers are not more distant from the wall than 2 lattice spacings.
We choose $\epsilon_w = 0.1$
since from studies of wetting phenomena in polymer mixtures~\cite{51}
we can safely expect that this choice leads to a wetting transition
temperature $T_w$ below the considered temperature.
Note that the transition temperature $T_c (D)$ of
the interface localization--delocalization transition occurs~\cite{30,31,32} 
is expected close to $T_w$, and we need $T > T_c (D)$ in order
to have the situation sketched in Fig.1a (interface delocalized and
on average in the middle of the thin film).
The absolute magnitude of both surface terms is chosen the same, the
sign opposite, so we realize  an antisymmetric wall situation, as
desired. 

The linear dimensions $L$ parallel to the walls are in the range from
$L = 64$ to $L = 512$, while $8 \le D \le 64$. 
Remembering that $R_g \approx 7$, 
our film thicknesses thus range from one $R_g$ to about $9 R_g$.

Since these system sizes are rather large, at least for the higher values
of $D$ and $L$, and interfacial fluctuations are notorious slow~\cite{31},
one must pay much attention to an adequate equilibration of the model system.
In the canonical ensemble the order parameter is conserved and we have
$\overline{\rho}_A = \overline{\rho}_B$.
Two different types of runs have been made:
In the first type of runs, ``slithering snake'' moves~\cite{47}
(a chain end of a randomly chosen chain is cut at one end and one attempts
to join it at the other end in a randomly chosen direction)
are mixed with the ``random hopping'' moves~\cite{47,48}
(one tries to move a randomly chosen monomer of a randomly chosen chain
in a randomly chosen lattice direction by one unit).
It is clear, however, that long range concentration fluctuations
will relax rather slowly with this algorithm --- diffusion of the chains
through the whole system is required.
To ease this problem, a second type of runs was tried where the two
types of moves mentioned above were combined with chain identity
exchanges.
To conserve the order parameter, one needs to transform a randomly chosen 
chain (1) from A to B and another randomly chosen chain (2) from B to A.
After some tests we found it useful to mix these moves
(slithering snake : random hopping : identity exchange) in the ratios
3:1:0.1 (note that the ``1'' for the random hopping means one attempted move 
per monomer of the chosen chain).
The total amount of these moves is then counted as ``4 Monte--Carlo steps
(MCS)'' in order to define a ``pseudotime'' unit for this mixed algorithm.
For the semi-grand-canonical ensemble, we always use an algorithm of this 
second type, the only distinction being that one attempts
an identity change of a single chain (A to B {\it or} B to A, respectively)
instead of an identity exchange between two chains.
Since the volume fraction is not conserved,
$\overline{\rho}_A = \overline{\rho}_B$ holds on average only, but
in general not for individual configurations generated.

Of course, a price one has to pay for these mixing of algorithms
is that one performs considerably less moves per second than
with algorithms that carry out reptation moves only.
E.g., on an IBM RISC 6000-250 workstation one needs only 30 sec to 
perform $10^6$ MCS per chain if only reptation moves are carried out, 
while the canonical moves including chain exchange take $t \approx 140$ sec,
the semi-grand-canonical moves even $t \approx 150$ sec.
In order to run a system with $L = 512$ and $D = 32$ for 
$10^6$ MCS for every chain in the system with semi-grand-canonical moves, 
700 hours CPU are needed.
Since we found it necessary to carry out runs that were partly up to 10 times
longer, and many combinations of $(L,D)$ need to be studied,
this project is not feasible at workstations as quoted.
Fortunately, more powerful machines
(including DEC 8400 and SG PowerChallenge) could be accessed.
\section{Theoretical background}
\label{s3}
\subsection{Definitions}	
\label{ss3.1}	
In this section we define the quantities that will be calculated
in the later sections and recall some of the basic theoretical
predictions.
We denote the coordinate of a point inside two walls as $(\vect{r},z)$
where $\vect{r} = (x,y)$ is a two-dimensional coordinate in the surface plane 
of the left wall (at $z = 0$), and $z$ measures the distance
from this wall.
From the local densities $\rho_A (\vect{r},z), \rho_B(\vect{r},z)$
of all monomers we define then a local order parameter $m(\vect{r},z)$ as
follows:
\beq
m(\vect{r},z) = 
\frac{\rho_A(\vect{r},z)-\rho_B(\vect{r},z)}
{\rho_A(\vect{r},z)+\rho_B(\vect{r},z)},
\label{eq2}
\eeq
while the average order parameter profile is
\beq
m(z) = L^{-2} \int \! d \vect{r} \; m(\vect{r},z).
\label{eq3}
\eeq
Here $\int \! d \vect{r}$ stands symbolically for a summation over the
$L^2$ lattice points in plane $z$, and $m(z)$ is defined at the discrete
points $z = 1,2,...,D-1$ only, of course.

A quantity of key interest is the order parameter correlation function
in the direction parallel to the walls,
\beq
g(r,z) = \frac{\langle m(\vect{r},z) \; m(0,z) \rangle 
- \langle m(0,z) \rangle^2}
{ \langle m(0,z)^2 \rangle - \langle m(0,z) \rangle^2},
\label{eq4}
\eeq
from which we can extract a correlation length $\xi_{||}$ by observing at 
large distances $r$ an exponential decay,
\beq
g(r,z=D/2) \propto \frac{1}{\sqrt{r}} \exp \left( - \frac{r}{\xi_{||}} \right).
\label{eq5}
\eeq
Here we have anticipated (as is also borne out by our numerical data
presented below) that the slowest decay of $g(r,z)$ occurs right in the 
center of the film, due to the interfacial fluctuations~\cite{30,31,32}.

We define the interfacial width $w$ from a fit to a $\tanh$ profile,
\beq
\rho_A(z) = \frac{1}{2} 
\left[ \rho_{A,coex}^{(1)} + \rho_{A,coex}^{(2)} \right]
+ \frac{1}{2} 
\left[ \rho_{A,coex}^{(2)} - \rho_{A,coex}^{(1)} \right] 
\tanh \left( \frac{z - D/2}{w} \right),
\label{eq6}
\eeq
where the densities of A monomers
at the coexistence curve $\rho_{A,coex}^{(1)},
\rho_{A,coex}^{(2)}$ are known from previous work~\cite{22,45}.
Alternatively, we can write Eq.\ref{eq6} in terms of $m(z)$ in a more
compact form
\beq
m(z) = m_b \tanh \left( \frac{z-D/2}{w} \right),
\label{eq7}
\eeq
with 
$m_b = [\rho_{A,coex}^{(2)} - \rho_{A,coex}^{(1)}]/[[\rho_{A,coex}^{(2)} 
+ \rho_{A,coex}^{(1)}]$,
noting that for our symmetric mixture 
$\rho_{B,coex}^{(2)} = \rho_{A,coex}^{(1)},
\rho_{B,coex}^{(1)} = \rho_{A,coex}^{(2)}$.
Actually, for our choice of $\epsilon = 0.03$ the
mixture in the bulk is essentially fully segregated~\cite{22},
$\rho_{A,coex}^{(1)} = \rho_{B,coex}^{(2)} \approx 0$,
$\rho_{A,coex}^{(2)} = \rho_{B,coex}^{(1)} \approx \rho_b$,
$m_b \approx 1$, and hence Eq.\ref{eq6} simplifies to
\beq
\rho_A (z) = \frac{1}{2} \rho_b \left[ 1 + 
\tanh \left( \frac{z-D/2}{w} \right) \right].
\label{eq8}
\eeq
Since we have chosen to work with only even values for $D$,
the ``Gibbs dividing surface''~\cite{1,2} at $z = D/2$ always 
coincides with a position at the lattice.

The fact that the bulk order parameter $m_b \approx 1$ in our model implies 
that in the present case the bulk order parameter fluctuations are
negligible, and any interplay of bulk and interfacial fluctuations also can 
be disregarded here.
In this sense, the present model is simpler than the Ising model considered 
in Refs.\cite{31,32}. (One cannot go to such low temperatures with 
$m_b \approx 1$ there because the interface roughening transition~\cite{52} 
intervenes.)

\subsection{Capillary waves}
\label{ss3.2}
Our choice of the $\tanh$ profile in Eqs.\ref{eq6}-\ref{eq8} is motivated 
by the fact that this form is predicted by mean field type theories, such 
as Cahn--Hilliard type~\cite{6} long wavelength theories~\cite{7,8} and the 
self-consistent field theory in the limit~\cite{12,13} $\chi N \to \infty$. 
All these theories, however, neglect fluctuations in the position of the 
``local'' Gibbs dividing surface which we may define as
[$\rho_{A,B}(x,y) = D^{-1} \int \! d z \; \rho_{A,B} (x,y,z)$]
\beq
z_{int} (x,y) = \frac{\rho_A (x,y)}{\rho_A (x,y) + \rho_B (x,y)} \; D
\label{eq9}
\eeq
It then is convenient to consider the deviation 
$h(x,y) = z_{int} (x,y) - D/2$ 
of this fluctuating interface from its average location.
As is well known~\cite{1,2,3,4}, the idea of capillary wave theory is 
to put the free energy cost of these fluctuations proportional to the 
increase in interfacial area caused by these fluctuations.
Assuming that the gradients 
$|\partial h / \partial x|,|\partial h / \partial y|$ 
are small, this yields (normalizing again by a factor $\kt$)
\beq
\Delta F_{CW} = \frac{\sigma}{2} \int \! dx dy \left[ 
\left( \frac{\partial h}{\partial x}\right)^2 +
\left( \frac{\partial h}{\partial y}\right)^2 \right],
\label{eq10}
\eeq
where it is assumed that $\sigma$ is simply the bulk interfacial tension 
(normalized per temperature).
By Fourier transformation one obtains from Eq.\ref{eq10} a simple 
Gaussian Hamiltonian, $h(\vect{q})$ denoting the Fourier component 
of $h(x,y)$ for wavevector $\vect{q}$,
\beq
\Delta F_{CW} = \frac{\sigma}{2} \sum_{\vect{q}} q^2 | h (\vect{q}) |^2,
\label{eq11}
\eeq
and from the equipartition theorem one can immediately conclude that the 
mean square value of $h(\vect{q})$ is~\cite{2}
\beq
\langle | h (\vect{q}) |^2 \rangle = \frac{1}{\sigma q^2}.
\label{eq12}
\eeq
This power law spectrum clearly leads to divergences when one computes
the local mean square displacement of the interface~\cite{1,2,3,4},
\bea
s^2 & \equiv & \langle h^2 (x,y) \rangle \nonumber \\
& = & \sum_{\vect{q}} \langle | h (\vect{q}) |^2 \rangle \nonumber \\
& = & \frac{1}{4 \pi^2} \int \! d \vect{q} \; \langle |h(\vect{q})|^2 \rangle \nonumber \\
& = & \frac{1}{2 \pi \sigma} \int \! dq \; \frac{1}{q} \nonumber \\
& = & \frac{1}{2 \pi \sigma} \ln \left( \frac{q_{max}}{q_{min}} \right)  
\label{eq13}
\eea
Since the integral $\int \! dq/q$ diverges logarithmically both for
$q \to 0$ and $q \to \infty$, we have heuristically used both a lower
cut-off $q_{min}$ and an upper cut-off $q_{max}$.
It is clear that the naive approximation leaving the $1/q^2$ spectrum 
in Eq.\ref{eq12} completely unmodified in between $q_{min}$ and $q_{max}$
and using these sharp cut-offs needs a closer investigation.
It could be necessary to cut off these divergences  in a smooth way,
using suitable correction terms in the capillary wave Hamiltonian,
Eqs.\ref{eq10}~and \ref{eq11}, in order to obtain a more accurate 
description~\cite{53}.
Even if one accepts Eq.\ref{eq13}, the correct choice of $q_{min}$ 
and $q_{max}$ is a problem, particularly for a polymer mixture.
For large enough $D$ such that no confinement effects are felt by these 
interfacial fluctuations, the minimal value of $q$ possible in the 
geometries of Fig.1a,b clearly is $q_{min} = 2 \pi / L$ (remember that 
we work with periodic boundary conditions), but the choice of $q_{max}$ 
is much more questionable: For a small molecule system away from the critical 
region, one just takes $q_{max} = 2 \pi / a$, $a$ being a molecular diameter, 
while near the critical point $q_{max} = 2 \pi / \xi$
as the correlation length $\xi$ then is the only important length scale in the 
system. For polymer mixtures in the strong segregation limit, we have three 
length scales to consider: the length scale $b$ of an effective bond, the 
radius of gyration $R_g = b \sqrt{N / 6}$, and the ``intrinsic'' interfacial 
width $w_0$, which is controlled by the Flory--Huggins parameter $\chi$. 
According to Helfand's~\cite{12,13} self-consistent field theory, this 
intrinsic width is
\beq
w_0^{SSL} = \frac{b}{\sqrt{6 \chi}}.
\label{eq14}
\eeq
However, the finding of recent computer simulations~\cite{22} that the $\chi$ 
parameter itself is not constant in the interfacial region in the strong 
segregation limit --- due to subtle rearrangements in the chain configuration 
(enhancement of self-contacts, etc.) --- has raised doubts on the accuracy of 
Eq.\ref{eq14}, and in fact Eq.\ref{eq14} is not in agreement with the 
interfacial widths observed in the simulations~\cite{21,22}.
Thus, the proper estimation of $w_0$ may be a problem.
Semenov~\cite{19} has suggested that $w_0$ should be used in $q_{max}$ 
and he concluded
\beq
q_{max} = \frac{2}{w_0}.
\label{eq15}
\eeq
Note the absence of a factor $\pi$ in Eq.\ref{eq15}:
While in $q_{min} = 2 \pi / L$ the constant $2 \pi$ simply is there due 
to the periodic boundary condition, there is no condition that actually 
would fix the constant of proportionality between $q_{max}$ and $w_0$~!

\subsection{Capillary wave broadening of the intrinsic profile: 
the convolution approximation}
\label{ss3.3}
The treatment of the previous subsection has implied that the local 
interface is defined by the coordinate $z_{int} (x,y) = D/2 + h(x,y)$ 
only, i.e., the interface is considered as a sharp boundary without 
any intrinsic structure.
While this ``sharp kink'' approximation is reasonable when long wavelength 
properties are concerned, it clearly does not make sense on small length
scales of order $w_0$: For length scales $L$ of order $2 \pi / q_{max}$ 
we would have $q_{max} / q_{min} = 1$, i.e., Eq.\ref{eq13} predicts a 
vanishing mean square interfacial width, which is not a physically 
sensible result.
The standard remedy~\cite{4} of this situation is to convolute the 
intrinsic profile $\rho_A^{(int)}(z)$ (which may be taken of the form of 
Eq.\ref{eq8} in our case) with a Gaussian distribution of local interface 
heights,
\beq
P(h) = \frac{1}{\sqrt{2 \pi s^2}} \exp \left( - \frac{h^2}{2 s^2} \right),
\label{eq16}
\eeq
such that the apparent profile becomes 
\beq
\rho_A^{(app)}(z) = \int_{-\infty}^{+\infty} \! 
dh \; \rho_A^{(int)} (z-h) P(h).
\label{eq17}
\eeq
For explicit calculations it is convenient to replace the $\tanh$ function 
by an error function with the same slope at the midpoint of the profile,
\beq
\rho_A^{(int)}(z) = 
\frac{1}{2} \rho_b \left[ 1 + \tanh \left( \frac{z-D/2}{w_0} \right) \right]
\approx \frac{1}{2} \rho_b \left[ 1 + \erf \left( 
\frac{\sqrt{\pi} (z-D/2)}{2 w_0} \right) \right]
\label{eq18}
\eeq
We now define the interfacial width $w$ in terms of the maximum slope of the 
apparent profile,
\beq
(2 w)^{-1}
\equiv \left. \frac{1}{\rho_b}
\frac{d}{dz} \rho_A^{app} (z) \right|_{z=D/2}.
\label{eq19}
\eeq
Using Eq.\ref{eq18} in Eq.\ref{eq17} we find
\beq
w^2 = w_0^2 + \frac{\pi}{2} s^2
\label{eq20}
\eeq
and combining this result with Eq.\ref{eq13} yields
\beq
w^2 = w_0^2 + \frac{1}{4 \sigma} \ln \left( \frac{q_{max}}{q_{min}} \right) .
\label{eq21}
\eeq
Eq.\ref{eq21} is the well known result that the mean square 
broadening due to capillary waves simply has to be added to 
the square of the intrinsic width.

\subsection{Delocalized interfaces in confined geometry}
\label{ss3.4}
While Secs.\ref{ss3.2},\ref{ss3.3} describe the situation 
of Fig.1a in the limit $D \to \infty$ keeping $L$ fixed, we 
now consider the inverse limit, keeping $D$ fixed but letting $L \to \infty$.
Then the lower cut-off is no longer $2 \pi / L$, but rather 
a correlation length $\xi_{||}$ comes into play~\cite{30,32}.
We briefly review the derivation of that length~\cite{32}.
One again uses a Hamiltonian of the form of Eq.\ref{eq10}, 
but now we need to explicitly consider potentials $V (h)$ 
exerted by the walls on the fluctuating 
interface~\cite{23,24,25,26,27,28,29,30,32}
\beq
{\cal H}_{eff} \{ h \} = \int \! dx dy 
\left\{ \frac{\sigma}{2} 
\left[ 
\left( \frac{\partial h}{\partial x} \right)^2 +
\left( \frac{\partial h}{\partial y} \right)^2 
\right] + V(h) 
\right\} 
\label{eq22a}
\eeq
or more explicitly
\bea
{\cal H}_{eff} \{ h \} & = & \int \! dx dy 
\left\{ \frac{\sigma}{2} 
\left[ 
\left( \frac{\partial h}{\partial x} \right)^2 +
\left( \frac{\partial h}{\partial y} \right)^2 
\right] \right. \nonumber \\
& & + 2 a_0 
\left( \frac{T-T_w}{T_w} \right) \exp 
\left(- \kappa D / 2 \right)
\cosh 
\left( \kappa h \right) \nonumber \\
& & \left. + 2 b_0 \exp 
\left( - \kappa D \right) \cosh 
\left( 2 \kappa h \right) 
\right\}
\label{eq22b}
\eea
Note that potentials of the form $\exp (-\kappa D / 2) \cosh (\kappa h)$
simply arise from exponentially decaying forces due to the walls
\[
\exp \left[ - \kappa z_{int} \right] +
\exp \left[ - \kappa (D-z_{int}) \right] =
2 \exp \left[ - \kappa D / 2 \right] \cosh \left( \kappa h \right).
\]
Here the first potential changes its sign at a temperature $T_w$ to 
make allowance for the fact that in the semi-infinite system a wetting 
transition occurs at $T_w$.
The constant $\kappa$ in mean field theory~\cite{30} is identified with 
the inverse bulk correlation length $\xi^{-1}$, while more refined 
treatments imply~\cite{30,54,55,56,57} that $\kappa^{-1}$ gets 
renormalized by a factor $(1+\omega/2)$ where $\omega$ is the 
famous parameter entering the theory of critical 
wetting~\cite{25,58,59,60,61},
\beq
\kappa^{-1} = \xi (1+\omega/2), \mbox{\hspace{1cm}} 
\omega = (4 \pi \xi^2 \sigma)^{-1} 
\label{eq23}
\eeq
Finally, $a_0,b_0$ are phenomenological constants that can be 
calculated from a microscopic theory of interfaces near walls 
for Ising models~\cite{62,63,64}, but are not known for polymer 
mixtures, of course.

Also the temperature dependence of the constant $\omega$ is known 
rather  accurately for the Ising model~\cite{65}, but not here.
Fig.2 shows a log-log plot of the expected behavior of
$\omega$ vs $N(\chi/\chi_{crit}-1)$.
In the mean field critical regime of a polymer mixture 
(i.e., $|1-\chi/\chi_{crit}| \ll 1$, but~\cite{50,66,67} 
$N |1 - \chi / \chi_{crit}| \gg 1$) one predicts~\cite{8,50}
for a lattice model where every site is taken by either an A~monomer
or a B~monomer (i.e, $\rho b^{-3} = 1$),
\beq
\xi = \frac{b}{6} \sqrt{N} \;
\left( 1-\frac{\chi_{crit}}{\chi} \right)^{-1/2},
\mbox{\hspace{1cm}} \sigma = \frac{8}{3 b^2 \sqrt{N}} 
\left(1 - \frac{\chi_{crit}}{\chi} \right)^{3/2}
\label{eq24}
\eeq
and hence the mean field prediction for $\omega$ is~\cite{68}
\beq
\omega_{MF} = 
\frac{27}{8 \pi}
\left[ N \left(1-\frac{\chi_{crit}}{\chi} \right) \right]^{-1/2},
\mbox{\hspace{1cm}} N \left| 1- \frac{\chi_{crit}}{\chi} \right| \gg 1.
\label{eq25}
\eeq
Very close to the critical point (i.e., for $N |1-\chi/\chi_{crit}| \ll 1$) 
one expects a crossover to the universal value of the Ising 
model~\cite{65} $\omega_{Ising} = 0.86$.
However, more relevant for the present simulations is the 
behavior of $\omega$ in the strong segregation limit.
While in the critical regime we have a simple relation 
between the (intrinsic) width of the interface $w_0$ and 
the correlation length~\cite{50}, $w_0 = 2 \xi$,
this is no longer true in the strong segregation limit 
(where $\xi \approx b \sqrt{N} / 6$ while $w_0$ is independent 
of $N$, cf. Eq.\ref{eq14}).
Since $\omega$ in the critical region also could be expressed 
as $\omega = [ \pi w_0^2 \sigma]^{-1}$,
we suggest that this expression
continues to be valid in the strong segregation limit rather 
than the expression given in Eq.\ref{eq23}, cf. Fig.2.
Using~\cite{12,13} $\sigma^{SSL} = b^{-2} \sqrt{\chi/6}$
and Eq.\ref{eq14} one obtains
\beq
\omega_{SSL} = \frac{6}{\pi} \sqrt{6 \chi}
= \frac{12}{\pi} \sqrt{\frac{3}{N} \frac{\chi}{\chi_{crit}}},
\label{eq26}
\eeq
where in the last step the mean field result $\chi_{crit} = 2 / N$ 
was used.
Consequently, we assume that also in the equation for $\kappa$  
(Eq.\ref{eq23}) one should replace $\xi^{-1}$ by $2 / w_0$
in the strong segregation limit.

Expanding $\cosh (\kappa h) \approx 1 + (\kappa h)^2 / 2$ and 
omitting constant terms, Eq.\ref{eq22b} reduces to
\bea
{\cal H}_{eff} \{ h \} 
& = & \int \! dx dy \left\{ \frac{\sigma}{2} \left[ 
\left( \frac{\partial h}{\partial x} \right)^2 +
\left( \frac{\partial h}{\partial y} \right)^2 \right] \right. \nonumber \\ 
& & \left. + \left[ a_0 \left( \frac{T-T_w}{T_w} \right) 
\exp \left(- \kappa D / 2 \right)
+ 4 b_0 \exp \left( - \kappa D \right) \right] \left( \kappa^2 h^2 \right)
\right\}
\label{eq27}
\eea
We see that the second term changes sign at~\cite{30,32}
\beq
T_c (D) = T_w 
\left[ 1 - 4 \; \frac{b_0}{a_0} \; \exp \left( - \frac{\kappa D}{2} \right) \right],
\label{eq28}
\eeq
which is the critical temperature of the interface 
localization--delocalization transition~\cite{30,32}.
Of course, even on the mean field level the approximation 
Eq.\ref{eq27} is only valid for $T  > T_c(D)$.
In this regime, Eq.\ref{eq27} is formally identical to 
Ginzburg--Landau theories of second--order phase 
transitions~\cite{1,2,3,4,50}, and thus it is straight forward 
to obtain the correlation length $\xi_{||}$, 
using the standard relation
\beq
\xi_{||}^{-2} = \frac{1}{\sigma} \left( \frac{\partial^2 V(h)}{\partial h^2}
\right)_{h=0}
\label{eq29}
\eeq
to obtain
\beq
\xi_{||}^{-2} = 
\frac{2 a_0}{\sigma} \; \kappa^{2} \; \exp \left( - \frac{\kappa D}{2} \right)
\left( \frac{T-T_c(D)}{T_w} \right) .
\label{eq30}
\eeq
As expected, $\xi_{||}$ shows a mean field type divergence 
at $T = T_c (D)$. Even for $T \gg T_c (D)$, however, $\xi_{||}$ is 
very large for high values of $D$ , due to the factor 
$\exp ( \kappa D / 4 )$ in the relation 
$\xi_{||} \approx \kappa^{-1} \sqrt{\sigma T_w / (2 a_0 T)} 
\exp ( \kappa D / 4 )$. 
If we now assume that it is the length $\xi_{||}$ rather 
than $L$ that cuts off the capillary wave spectrum, we conclude 
$q_{min} = 2 \pi / \xi_{||}$ and hence Eq.\ref{eq21} becomes
\beq
w^2 = w_0^2 + \frac{\kappa D}{16 \sigma} + const,
\label{eq31}
\eeq
the constant being 
$const = (4 \sigma)^{-1} \ln 
\left[ \kappa^{-1} q_{max} \sqrt{\sigma T_w / ( 2 a_0 T)} / (2 \pi) \right]$.
As discussed above, this constant is hard to estimate, 
and in the simplest approximation it has been neglected~\cite{33}.
This is justified in the mean field critical region, 
since $\ln[...]$ should be of order unity, and 
$w_0^2 \gg (4 \sigma)^{-1}$ there, but it is not 
obvious that this approximation is accurate in the 
strong segregation limit. And again the question 
must be asked how reliable it is to use $\xi_{||}$ 
from Eq.\ref{eq30} as a cut-off in a calculation 
with the ``free'' capillary wave Hamiltonian, 
Eq.\ref{eq10}, rather than calculating $s^2$ 
directly from the full Hamiltonian, Eq.\ref{eq22b}, 
without further approximations. 
Also effects such as a possible position-dependence of the
interfacial stiffness (the prefactor of the term
$(\partial h / \partial x)^2 + (\partial h / \partial y)^2$)
are neglected, although theories have suggested such effects
for Ising-like systems~\cite{62,63,64}
Hence a test of 
Eq.\ref{eq31} by computer simulation is of significant interest.

\section{Interfacial profiles, interfacial widths, and the correlation 
length $\xi_{||}$}
\label{s4}
Fig.3 shows some raw data of order parameter profiles in a semi-grand-canonical
simulation using $L = 256$ for films of various thicknesses $D$ ranging from
$D = 16$ to $D = 64$.
The broadening of the profiles with increasing 
film thickness $D$ is clearly recognized, and qualitatively these 
data are very similar to earlier results for the Ising model~\cite{33}.
From a fit of these data to $\tanh$ profiles, estimates for the 
apparent width $w$ as function of $L$ and $D$ is obtained.

In Fig.4, this width $w$ is plotted vs $D$ for $32 \le L \le 512$
in the semi-grand-canonical ensemble (sg) 
and the canonical ensemble with chain exchange (e), respectively. 
We find three different regimes in our simulations.
In the first one, $D$ is small, and the data depend neither on $L$ nor on 
the type of chosen ensemble. 
The increase of $w^2$ with $D$ for $D \le 16$ 
even is somewhat stronger than the expected (Eq.\ref{eq31}) linear variation.
This is due to the fact that for very small $D$ ($D < w_0$) not only
capillary wave type fluctuations are suppressed, but even the 
intrinsic profile gets ``squeezed'', and we expect $w \propto D$ 
(or $w^2 \propto D^2$) as $ w \to 0$; there is just an essentially 
linear interfacial profile between such walls at extreme proximity 
of each other.
In particular, for $ D \to 0 $ the interfacial width $w$ must vanish.
Note that additional effects can also be expected due to the 
squeezing of the chains in quasi-two-dimensional configurations 
if $D \le R_g$.

In the second regime for $24 \le D \le 56$ and large enough $L$, 
data in the semi-grand-canonical ensemble do show the expected variation 
$w^2 \propto D + const$.
For small $L$, however, a crossover to the third regime with a steeper rise 
occurs.
This crossover is clearly seen for $L \le 128$.
We expect that for $L = 256$ a similar upturn should occur for 
$D = 64$, but this is not observed due to the 
relative large statistical error of this last point 
(note that the exponential increase of the correlation 
length $\xi_{||}$ with thickness $D$, Eq.\ref{eq30}, 
leads to a dramatic slowing down of all 
Monte--Carlo simulations, and this is responsible for the 
fact that a small increase of $D$ causes a large 
increase of statistical errors).
This upturn of the curves for large $D$ and relatively 
small $L$ is interpreted by noticing that we now enter 
a regime where $L < \xi_{||}$, and then Eq.\ref{eq31} 
is not valid: In a semi-grand-canonical simulation with 
periodic boundary conditions the dominating fluctuation 
is a mode with wavenumber $q = 0$, i.e., a uniform 
displacement of the interface as a whole~! Thus, there is a 
finite fraction of the thickness $D$ where the interface can 
wander back and forth like a rigid straight interface, without 
feeling much effects of the walls.
In the limiting case of $D \to \infty$ and $L$ small,
the interface is freely fluctuating, and thus Eq.\ref{eq19} 
gives a linear dependence $w = D/2$~\cite{ftnt}. 

In the canonical ensemble, on the other hand, 
this third regime is very different.
For $L < \xi_{||}$ no significant further 
increase of $w^2$ with $D$ occurs; rather $w^2$ then 
saturates at a plateau value because in the canonical 
ensemble a fluctuation of the mean interface position 
is suppressed, and the capillary wave spectrum now is again 
limited by $L$ ($q_{min} = 2 \pi / L$) rather than $\xi_{||}$.
Hence, instead of Eq.\ref{eq31} we now have from Eq.\ref{eq21}
\beq
w^2 = w_0^2 + \frac{1}{4 \sigma} \ln L + \frac{1}{4 \sigma}
\ln \left( \frac{q_{max}}{2 \pi} \right)
\label{eq32}
\eeq
independent of $D$, which is the standard logarithmic 
variation of the interface squared width with the lateral 
linear  dimension, as it has often been discussed in the 
literature~\cite{1,2,3,4,22,41,41b}.
Considering the values of $w^2$ for $D = 64$
and $L = 32$, $64$, and $128$ as independent from the finite thickness,
we can extract from the slope of $w^2$ vs $\log L$ the interfacial
tension $\sigma$ and find very good agreement with the simulations
of M\"uller et al~\cite{22}, i.e., $\sigma = 0.015$.

In order to bring out more clearly these three regimes 
for the semi-grand-canonical ensemble,
we compare in Fig.5a the data for $L = 256$ (which cover 
the most extensive range of $D$ and are least affected by the finite size 
effects associated with the smallness of $L$) with theoretical prediction, 
Eq.\ref{eq31}.
As pointed out in Sec.\ref{ss3.4}, neither the choices of $w_0$ nor the 
choice of $\kappa$ nor the choice of the additive constant is unique.
But the choice made for the parameters in Fig.5a 
--- $\xi = 3.6$ from Monte--Carlo simulations and
$ w_0 = w_0^{SCF} = 4.65 $ as determined in a self-consistent field 
calculation for our model~\cite{21} --- is at least plausible,
even if the good agreement with the data for $D \ge 40$ is a mere lucky 
accident.
In Fig.5b, a linear plot of $w$ vs $D$ checks for both linear regimes:
In the first regime for small $D$, we find one straight line through the 
origin, including data where $w \le w_0$ and using $w(D=0) = 0$.
On the other hand, the crossover to the third regime for large $D$
can be seen whenever $\xi_{||}$ becomes comparable to $L /2$.
Since $\xi_{||}$ depends on $D$ as $\xi_{||} \propto \exp (\kappa D / 4)$
(Eq.\ref{eq30}), it is clear that the location of the onset of the 
crossover shifts to larger $D$ with increasing $L$ while the first crossover
(when $w$ reaches its ``intrinsic'' value) is independent of $L$, of course.
When $L$ is very small ($L = 32$), the second regime 
(where Eq.\ref{eq31} holds) is completely suppressed.
For $L \ge 64$, this regime becomes more and more pronounced and
an increasing number of data points fall on the expected $L \to \infty$ 
curve.
Fig.5 illustrates how difficult it is to give interfacial widths observed in
simulations a precise meaning.

We now discuss the correlation function $g(r,z)$ (Eq.\ref{eq4}), see Figs.6-8.
In Fig.6, an example for $L = 256$ and $D = 32$ in the semi-grand-canonical
ensemble is shown.
It is seen that near walls this quantity has a few 
oscillations for small $r$,
reflecting the structure of the radial density distribution function,
while in the center of the thin film ($z = 16$ in Fig.6) a rather
slow decay occurs.
This slow decay is the hallmark of the ``soft mode phase'' described by
the theory of Sec.\ref{ss3.4} resulting from interfacial fluctuations.
In Figs.7~and~8, the correlation function $g(r) \equiv g(r,z=D/2)$ 
in the center of the film is studied.
A reasonable fit to the expected form of the correlation
function (Eq.\ref{eq5}) is indeed possible.
This is shown in Fig.7 for different film thicknesses ($16 \le D \le 48$)
using $L = 512$ and the semi-grand-canonical ensemble.
The data points (symbols) are fitted to Eq.\ref{eq5} 
(solid lines), and in this way
the correlation length $\xi_{||}$ is determined.
However, we must add the caveat that correlation functions of the order of
$10^{-3}$ and distances of the order of $r \ge 40$ are clearly very hard to
measure reliably in computer simulations.
From Fig.8 it is clear that simulations in the canonical ensemble underestimate
$g(r)$ at large $r$ systematically if they are based on local moves and 
slithering snake moves only.
Such observations as shown in Fig.8 have led us to use the
present canonical algorithm where also two identity exchanges of chains
(A \ra B and B \ra A) are attempted as an additional move (data shown by
squares in Fig.8).
It is also disturbing, however, that there seem to be small systematic
differences between the data for the canonical and the semi-grand-canonical
ensemble (the estimates for $\xi_{||}$ from Fig.8 are $\xi_{||} (sg) = 22.5$
(circles) and $\xi_{||}(e) = 20.0$ (triangles), even though $L = 256$ exceeds 
$\xi_{||}$ by about an order of magnitude.
Clearly, for $L \to \infty$ both ensembles should yield identical results
for $\xi_{||}$, and in fact for $L = 512$ and $D \le 40$ this is found
(cf. Fig.9).
Presumably the observed discrepancy tells that the errors in $\xi_{||}$
(due to statistical errors of $g(r)$) are of order of $10 \%$,
or there are unexpectedly large finite size corrections.

Nevertheless it is gratifying that the data for $\xi_{||}$ are compatible
with the predicted exponential variation 
$\xi_{||} \propto \exp ( \kappa D / 4)$,
as shown in Fig.9. The same choice of parameters as assumed in Fig.5a for the 
interfacial width (Eq.\ref{eq31}) gives a good representation of $\xi_{||}$
as function of $D$.
This, in turn, implies that a lot of the ambiguities about the choice of
parameters is avoided if we interpret Eqs.\ref{eq21}~and~\ref{eq31} in the form
$w^2 = w_0^2 + (4 \sigma)^{-1} \ln (q_{max} \xi_{||} / 2\pi)$
and simply use the measured $\xi_{||}$ in this equation:
a straight line with the same slope as included in Fig.5a must result,
irrespective of what one assumes for $w_0$, $q_{max}$, $\kappa^{-1}$,
and $\omega$.

\section{A direct test of capillary wave concepts for interfacial fluctuations}
\label{s5}
The good agreement between the simulation results shown in Figs.4,5,9 and
theoretical concepts based on the capillary wave Hamiltonian suggests to 
study the interfacial fluctuations in more detail.
In order to do this, it is convenient to split the system up into columns of
block size $B \times B$ and length $D$ and to determine in each column
a local interface position $h(x,y)$ using the concept of the Gibbs dividing
surface (in a constant region in the center of the film in order not to
be affected by bulk fluctuations).
Fig.10 shows snapshot pictures of instantaneous configurations of the interface
$h(x,y)$ obtained in this way for a minimal coarse--graining 
$B = 2$ (columns have the width of the size of one monomer) and the 
physically more plausible choice $B = 8$ (comparable to the radius of 
gyration).
While for $B = 2$ an extremely rugged ``interfacial landscape'' results,
for $B = 8$ rather pronounced large-amplitude interfacial fluctuations
are clearly visible.

Having found $h(x,y)$, we can proceed in the spirit of Sec.\ref{ss3.2} and
decompose $h(x,y)$ into Fourier modes $h(q)$, taking into account
that possible $q$-vectors on the lattice are given as 
$\vect{q} = (q_x,q_y) = 2\pi/L \; (i,j)$ with $i,j$ integers.
For simplicity, we consider only a wavevector in a lattice direction, and hence
classify the $h(\vect{q})$ as $h_i$.
Since Eq.\ref{eq12} implies $\langle | h_i |^2 
\rangle = (\sigma q_i^2)^{-1} \propto
i^{-2}$, we present a $\log-\log$ plot of
$\langle | h_i |^2 \rangle$ vs $i^{-2}$ in Fig.11.
Pure capillary waves in this representation should show up as straight 
lines with a slope of unity.
For $L = 256$ and $D = 56$ and $64$, this is indeed found for 
$2 \le i \; (< 20)$, but even for this largest studied thicknesses we still
see deviations due to finite $D$ for $i  = 1$~!
For $D \le 48$, this effect becomes more important and modes $i = 2,3, ...$
are affected by the finite thickness $D$. 
One could suspect that there is a problem of equilibration
of the largest
length scales of interfacial fluctuations, but this is not a 
problem here
as an analysis of the autocorrelation function 
$C_{hh} (t) = [ \langle h_i(t) h_i(0) \rangle - \langle h_i \rangle^2 ] / 
[ \langle h_i^2 \rangle - \langle h_i \rangle^2 ]$
of the modes as a function of time measured in Monte--Carlo steps
(MCS) shows (Fig.12).
The relaxation times of these fluctuations 
($\tau = {\cal O} (10^4 MCS)$) are much smaller than the
simulation runs ($\tau = {\cal O} (10^6 MCS)$).

An alternative analysis is to vary the block size $B$ over a wide range
and estimate the mean square width $w^2$ as a function of block size
This is shown in Fig.13 for $L = 128$.
Here the interfacial position $h$ is determined in all columns. The monomer
profile is taken relative to this position, $\rho_A (z-h)$, and after
averaging over all columns in the system, the interfacial width is determined
by a fit to a $\tanh$ profile.
For $D = 56$ and $64$, 
one sees a constant slope for $B \ge 8$, as expected for
capillary waves independent of film thickness $D$.
Note that the largest block size $B = 128$ is different from the smaller ones.
For $B = L$ periodic boundary conditions apply. 
Thus, the capillary wave spectrum is cut off
by $q_{min} = 2 \pi / L$.
In contrast, for $B < L$ the boundary conditions are free (as in
experimental situations)
and a supplementary mode appears characterized by $q = \pi / B$.
For $D = 16$, there is an upper cut-off block size
which gives a $w$ independent of $B$, but typical for $D = 16$.
For $24 \le D \le 48$, we find a rather slow 
crossover from a $D$--dominated to a 
$L$--dominated interfacial width for the largest block size.
On the other hand, the curves differ as well for all the small block sizes
in function of the film thickness $D$, and this difference is almost
constant for $4 \le B \le 32$.
This is a sign that there are two contributions to the variation
of the interfacial width with $B$ and $D$:
The first arises from capillary waves; it depends on the block size $B$,
and the finite thickness $D$ acts as a new cut-off length scale $\xi_{||}$.
On the other hand, there is a second contribution 
that depends on $D$ only, not on $B$.
It can thus be attributed to a change in the ``intrinsic width'' 
of the interface due to the presence of the walls.

For very small block sizes, $w^2$ becomes flat, and from this crossover one 
hence might attempt to estimate both $w_0$ and $q_{max}$ directly.
This would yield $w_0 \approx 3$ and $q_{max} \approx 2 \pi / 5$ for 
$D = 24$, and $w_0 \approx 4$ and $q_{max} \approx 2 \pi / 6$
for $D = 64$.
But exact values are hard to give since there is a no sharp
cut-off, but rather a smooth crossover.
The three important length scales 
monomer size, correlation length $\xi$, and 
width $w$ being close to each other,
this analysis does not really clarify the questions concerning the cut-offs.

Nevertheless the change of the intrinsic width with $D$ extracted from Fig.13
can be included quantitatively in our capillary wave analysis.
For this purpose, we define $w (D,B = 8) = w_0 (D)$ as the intrinsic width
of our system.
With this choice we are on the safe side, i.e. the further variation
of $w(D,B)$ in function of $B$ can be described in terms of a 
capillary wave broadening. For $B < 8$, Fig.13 indicates the onset of the
crossover to a saturation value of $w_0$.
We take then the data from Fig.5b for $w^2(D)$ and in Fig.14
$w_c^2 (D) = w^2 (D) - w_0^2 (D)$ is plotted.
In that way, a constant slope of $w_c^2$
vs $D$ is obtained over the whole range of simulations as required by
Eq.\ref{eq31}.
The straight line shows the theoretical prediction with the same choice of
parameter as before.
The main result is that both slopes agree very well.
Since in our case $w_0^{SCF} \approx w_0^{sim} (D \to \infty, B = 8)$,
even the absolute values are the same 
($w_0^{sim}$ denotes the interfacial width extracted from simulations).
Of course, from our data one could as well define $w (D,B=4)$ as intrinsic 
width.
Then we would recover the theoretical result by choosing a constant term 
in Eq.\ref{eq31} larger than zero.

Finally Fig.15 considers the distribution of interface locations $P(h)$.
It is seen that Gaussian distributions give a very good fit to such
data.
Using $w^2 (D,L = 128) = w_0^2 (D) + (\pi/2) s^2 (D)$ and 
$w^2 (B = 8)$ from Fig.13, we find 
$w^2 (D = 24,L = 128) = 15.1 + (\pi/2) \; 3.54^2 = 34.8$ and
$w^2 (D = 64,L = 128) = 21.8 + (\pi/2) \; 4.67^2 = 56.1$,
in good agreement with the direct results in Fig.13.
This shows once more the self-consistency of our analysis.

\section{Conclusion}
\label{s6}
In this paper we have studied the size-dependence of interfacial profiles
between coexisting phases of mixtures confined between parallel walls,
considering both the dependences on the linear dimension $L$ parallel to
the interface and the perpendicular linear dimension $D$.
The temperature of the system is above the wetting temperature $T_w$,
and the interface is delocalized and on average in the center of the film.
We argue that in this regime 
the interfacial correlation length 
$\xi_{||} \propto \exp (\kappa D / 4) $, 
proposed first by Parry and Evans~\cite{30} assuming
that $\kappa^{-1}$ is the correlation length $\xi$ of concentration
fluctuations in the bulk, acts like a cut-off for the capillary wave
spectrum, $q_{min} = 2 \pi / \xi_{||}$. Using the convolution of capillary
wave broadening with the ``intrinsic'' profile of width $w_0$,
we propose in the limit $L \gg \xi_{||}$
the formula for the total interfacial squared width
$w^2 = w_0^2 + (4 \sigma)^{-1} \ln ( q_{max} / q_{min} ) \approx
w_0^2 + (16 \sigma)^{-1} \kappa D$, where $\sigma$ is the interfacial 
tension.
This phenomenological approach is tested by a Monte--Carlo simulation of a
model for a symmetrical polymer mixture in the strong segregation limit,
and reasonable agreement is found, although the precise choice of parameters
(cut-off $q_{max}$, length $\kappa^{-1}$, capillary parameter $\omega$
in the relation $\kappa^{-1} = \xi (1+\omega/2)$ proposed by
recent theories~\cite{54,55,56,57}) may need refinement.

Of course, the above formula is only an asymptotic result for $D \gg w_0$;
for smaller values of $D$ we find a regime where capillary wave type
excitations are not yet important, and rather the intrinsic profile itself
is ``squeezed down'' by the confinement.
In this region $w \propto D$ (with a prefactor considerably smaller then 
unity) until $w$ has reached roughly the value $w_0$ of the intrinsic 
interface without confinement.
Note that there $D/2$ is comparable to the largest eigenvalue of the
gyration tensor of the chains, and hence an effect on the chain
configuration due to the confinement is already expected.
This effect will be studied more closely in future work.

The limit $\xi_{||} > L$ deserves attention, too.
We find that the above behavior $w^2 \approx w_0^2 + (16 \sigma)^{-1} \kappa D$
only holds until about $\xi_{||} = L/2$,
and then a regime sets in which depends on the type of ensemble used:
In the canonical ensemble, the average position of the interface is fixed 
at the
middle of the system, and the capillary wave spectrum is cut off by
$q_{min} = 2 \pi / L$ rather than $q_{min} = 2 \pi / \xi_{||}$.
As a result, $w^2 = w_0^2 + (4 \sigma)^{-1} \ln ( q_{max} L / 2 \pi)$
in this regime, independent of $D$.
In the semi-grand-canonical ensemble, on the other hand, there is 
also a fluctuation described by a wavenumber $q=0$, i.e., a uniform
mode in which the interface fluctuates back and forth as a whole
without bending.
This leads to a behavior $w \propto D$ for $D \to \infty$.

As a result, the meaning of interfacial profiles and interfacial widths is 
rather delicate.
This is not only a problem of computer simulations --- it is encouraging
to note that the analysis presented here is immediately useful for the
interpretation of corresponding experiments~\cite{33,42}.
Of course, in experiments it may be sometimes necessary to work
with long range surface forces (inverse power law decay) rather than
the exponential decay assumed here.
It is obvious that this will reduce the exponential growth of $\xi_{||}$
with $D$ into a power law growth as well, and since 
$\ln \xi_{||} = \ln D^p = p \ln D$ for any power $p$, we obtain
a behavior $w^2 \propto \ln D$ rather than $w^2 \propto D$,
as found here.
A detailed analysis of this situation is still lacking.

Many of our findings carry over to the other simulation geometries
considered in Fig.1, such as antiperiodic boundary conditions or fully
periodic boundary conditions.
Note that in the latter case only the canonical ensemble is useful
because otherwise the situation with two interfaces is unstable.
In the antiperiodic case both ensembles may be used, but since one
translates the interface to the center of the system by construction,
the ``uniform mode'' leading to the above behavior $w^2 \propto D^2$ is 
absent, and the finite size effects in the canonical and semi-grand-canonical
ensembles are very similar then.
We do hope that the present analysis will stimulate a careful assessment of 
these finite size effects in future simulations of interfacial properties
since we feel that much previous work has been hampered by inadequate
interpretation of fluctuation--broadened interfacial profiles as
``intrinsic profiles''.
The present work indicates that a unique characterization of ``intrinsic
profiles'' encounters difficulties even if size effects are carefully
taken care of.

\subsection*{Acknowledgements}
Support by the Deutsche Forschungsgemeinschaft (DFG), grant No.Bi-314/3-4,
and by the Bundesministerium f\"ur Bildung, Wissenschaft, Forschung
und Technologie (BMBF), grant No.03N8008C, is gratefully acknowledged.
We thank J.Klein and T.Kerle for stimulating discussions.
We acknowledge access to the computer facilities of the ZDV, 
University of Mainz, and the RHR Kaiserslautern.

%% file: figures.tex
\newpage
\section*{Figure captions}
\bd
\item[Fig.1:] Typical geometries used in computer simulations (a,b) and in
experiment (c) in order to study interfacial profiles.
Only one lateral direction parallel to the average interface plane is
shown.
The schematic plots indicate ``coarse--grained snapshots'' of the system,
i.e., bulk fluctuations in both the A-rich phase and the B-rich phase
are averaged out, and the resolution perpendicular to the interface
does not resolve the ``intrinsic'' interfacial profile in these pictures.
Case~(a) shows geometries where a single interface 
occurs~\cite{22,31,32,33,36}, while case~(b) refers to a situation with two
(on average parallel) interfaces confining a slab like region of B-rich
phase in between~\cite{37}.
While the left case of (a) has the disadvantage that perturbations due to the
two hard walls must affect the interfacial profile if $D$ is not large
enough, the right case of (a) is only possible for strictly symmetric mixtures
(``antiperiodic boundary conditions'' means that an A-particle that leaves
the simulation box on the right side reenters as a B-particle on the left side,
while no identity switches of particles occur across periodic boundaries).
In case (c) it is assumed that the direction of gravity is from left to
right and that the B-rich phase is favored by the air
while the A-rich phase is favored by the right wall (walls of the container
at the bottom and at the top are assumed to be neutral).
Note that here linear dimensions are not to scale (e.g. the sample lateral
dimension $L_{sample}$ may be macroscopically large and exceed the
lateral resolution length scale $L$ by several orders of magnitude), and
also the fluctuations of the coarse--grained interface are simplified
(short wavelengths being omitted, amplitudes of long wavelengths exaggerated
for clarity).
\item[Fig.2:] Log-log plot of $\omega$ vs $\chi / \chi_{crit} - 1$,
for chain lengths ranging from $N = 128$ to $N = 1024$,
showing the predictions for $\omega_{Ising}$, $\omega_{MF}$ (Eq.\ref{eq25}),
and $\omega_{SSL}$ (Eq.\ref{eq26}) using $\rho b^3 = 1$.
\item[Fig.3:] Order parameter profile $m(z)$ vs $z-D/2$ 
for films of thicknesses
$D = 16,32,48,$ and 64, as indicated by different symbols.
The interface is centered at the origin $ z = D/2 $,
walls being situated at $ z = 0 $ and $ z = D $.
All data refer to a lateral system size of $L = 256$ and the 
semi-grand-canonical ensemble.
The lines through the data points are fits to 
$m(z) = m_b \tanh \left[ (z-D/2) / w \right]$.
Note that statistical errors are smaller than the size of the symbols.
\item[Fig.4:] Plot of the square of the apparent interfacial width $w^2$
vs film thickness $D$ for several choices of $L$ ($L = 32, 64, 128, 256,$ 
and 512) and two choices of statistical ensembles: the canonical ensemble
including identity exchanges (opaque symbols marked by (e))
and the semi-grand-canonical ensemble (filled symbols marked by (sg)).
Note that the accuracy of these data is very good for small $D$, while
for the two largest values of $D$ ($D =56,64$) statistical errors are
estimated to exceed the size of the symbols several times.
For $L = 512$, values of $w^2$ are determined up to $D = 48$
and, within statistical error, no difference between the ensembles is
found and, thus, points in Fig.4 lie on top of each other.  
\item[Fig.5:] {\bf a)} Squared interfacial width $w^2$ vs film thickness $D$
for $L = 256$ in the semi-grand-canonical ensemble.
The straight line shows Eq.\ref{eq31}, neglecting the last term on the
right hand size and using $\kappa^{-1} = \xi ( 1+ \omega / 2)$,
$\xi = 3.6$, $\omega^{-1} = 4 \pi \xi^2 \sigma$, $\sigma = 0.015$,
and $w_0 = w_0^{SCF} = 4.65$~\cite{21}.\\
{\bf b)} Interfacial width $w$ plotted vs $D$ for $L = 32, 64, 128$, and
$L = 256$ using the data from the semi-grand-canonical ensemble simulations.
For small $D$, one has a linear variation $w \propto D$ up to $w \approx w_0$
(squeezed intrinsic interface).
Arrows show that the regime of free interface fluctuation sets in for
$\xi_{||} = L/2$.
The dashed line is the limiting case of completely free fluctuations of the
interface, i.e., $w = D/2$.
Also the self-consistent field estimate for the intrinsic interfacial width
($w_0^{SCF}$) is shown. 
Data for $\xi_{||}$ have been taken from Fig.9 below.
\item[Fig.6:] Correlation function $g(r,z)$ (Eq.\ref{eq4}) plotted as function
of both variables $r = \sqrt{x^2 + y^2}$ and $z$ for the case $L = 256$, 
$D = 32$.
\item[Fig.7:] Semilog plot of correlation functions 
$g(r) \equiv g(r,z=D/2)$ in the center of the thin film vs $r$ for $L = 512$
and general choices of $D$ as indicated.
All data refer to the semi-grand-canonical ensemble.
Curves are fits to $g(r)$ to the form
$g(r) \propto r^{-1/2} \exp ( -r/\xi_{||} )$, cf. Eq.\ref{eq5}.
\item[Fig.8:] Semilog plot of correlation functions $g(r)$ vs $r$ for
$L = 256$ and $D = 32$ for different ensembles:
semi-grand-canonical ensemble (circles), canonical ensemble (triangles) 
including chain identity exchanges, 
while squares indicate results with local moves and slithering snake 
moves only. Curves are fits to Eq.\ref{eq5}.
\item[Fig.9:] Correlation length $\xi_{||}$ plotted vs $D$ obtained by a fit
of $g(r)$ to Eq.\ref{eq5} for $L=512$ and film thicknesses $D \le 48$ using 
both data for the canonical ensemble (triangles) and the semi-grand-canonical
ensemble (circles).
Within statistical errors, both ensembles yield identical results for 
$D \le 40$, but differ for $D = 48$.
Full curve is the formula $\xi_{||} = \xi \exp (\kappa D / 4)$,
using $\kappa^{-1} = \xi ( 1+\omega/2 )$ and $\xi = 3.6$ as obtained from
an independent simulation.
\item[Fig.10:] Typical plot of a snapshot picture of the local interface
positions $h(x,y)$ for $D=64$, $L=64$ using the minimal coarse--graining
(i.e., block size $B=2$) in case (a) and modest coarse--graining ($B=8 \approx
R_g$) in case (b).
\item[Fig.11:] Log-log plot of the mean square Fourier component 
$\langle | h_i |^2 \rangle$ of the interfacial fluctuations vs $i^{-2}$
for $L = 256$ and the canonical ensemble.
Several choices of $D$ are included as indicated in the figure.
The straight line shows the theoretical prediction from the
capillary wave Hamiltonian, i.e.,
$\langle | h_i |^2 \rangle 
= \left( \sigma q^2 \right)^{-1} 
= \left[ L / (2 \pi i) \right]^2 \sigma^{-1}$.
\item[Fig.12:] Semilog plot of the autocorrelation function 
$C_{hh}(t) \equiv [ \langle h_i(t) h_i(0) \rangle - \langle h_i \rangle^2 ] / 
[ \langle h_i^2 \rangle - \langle h_i \rangle^2 ]$ vs time $t$ (in 
Monte--Carlo steps MCS). Data refer to $L= 256$, $D=48$ using the canonical
ensemble with chain identity exchanges.
The two slowest modes ($i=1,2$) are shown. Lines are fits of 
$- \ln [ C_{hh} (t) ]$ to an exponential decay.
Even for the slowest mode ($i=1$) the autocorrelation time $\tau \approx
1.5 \;10^4$ MCS is much smaller than the time of typical simulation runs.
\item[Fig.13:] Plot of the squared interfacial width $w^2$ vs the block
size $B$ (note the logarithmic scale) for a system of lateral linear
dimension $L = 128$ and thicknesses $D$ as indicated.
\item[Fig.14:] Same date as presented in Fig.5a 
($L = 256$, semi-grand-canonical ensemble), but only the contribution
$w_c^2 (D) = w^2 (D) - w_0^2 (D)$ due to capillary waves is shown. 
The intrinsic width $w_0 (D)$ is taken from Fig.13 
($L =128$, canonical ensemble) as the width determined
in blocks of $B = 8 \approx R_g$.
The straight line is the broadening according to theory (cf. Eq.\ref{eq31}),
i.e., $w_c^2 = \kappa D / (16 \sigma)$.
\item[Fig.15:] Distribution of interface locations $P(h)$ for $L=128$,
$24 \le D \le 64$ using a coarse--graining size $B = 8$.
Lines are fits to the theoretical assumption of a Gaussian distribution
$P(h) = (2 \pi s^2)^{-1/2} \exp [ - h^2 / (2 s^2) ]$.
The values of $s$ obtained by this fit are indicated.
\ed

\newpage
\pagestyle{empty}

\LARGE
\unitlength=1mm
\begin{picture}(150,150)
\put(0,0){
\psfig{figure=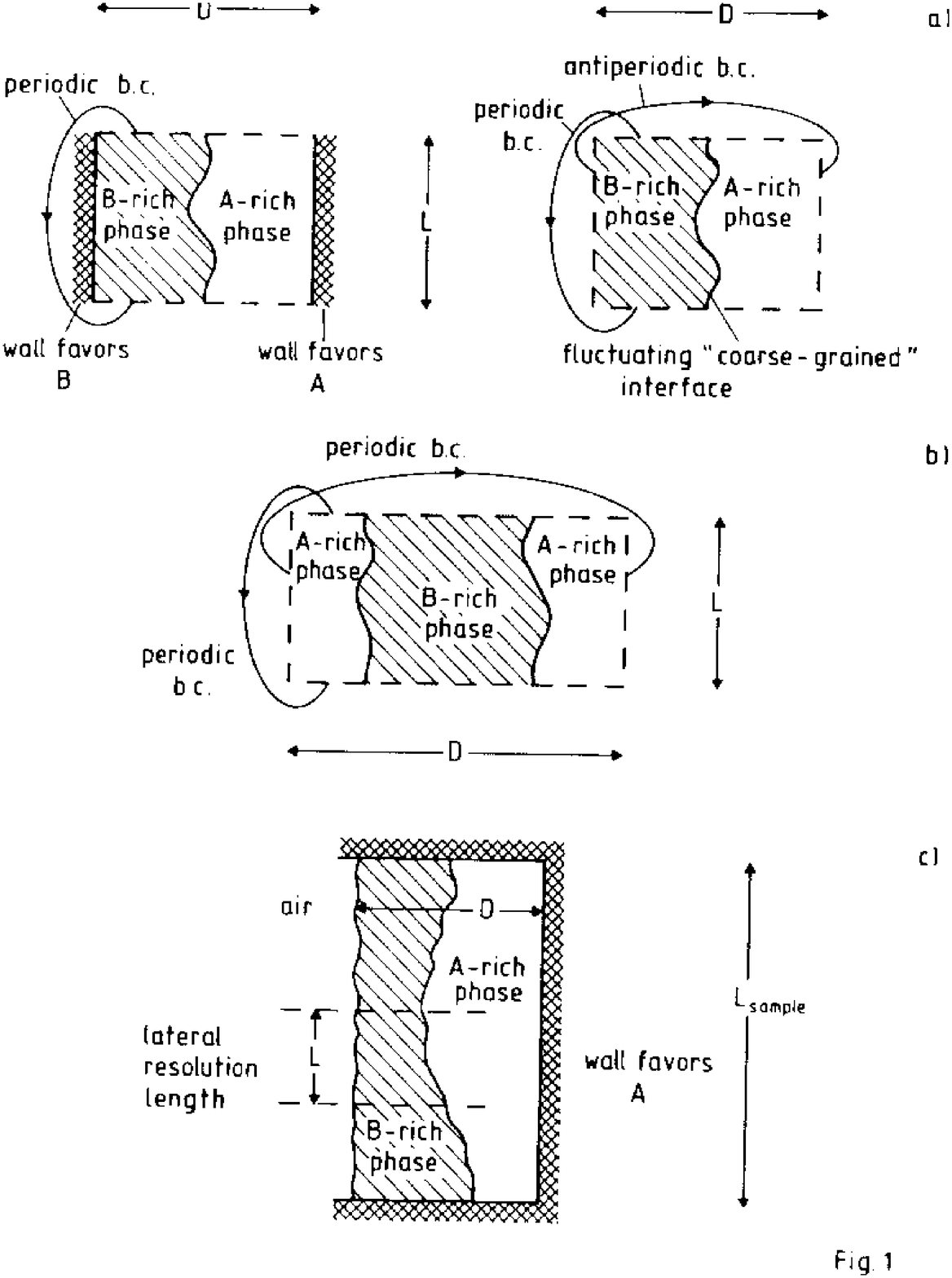,width=140mm,height=130mm,angle=1}
}
 
\end{picture}
\vfill
\normalsize
{\tt
\noindent
Figure 1\\
Werner et al\\ JCP
}

\newpage
\pagestyle{empty}

\LARGE
\unitlength=1mm
\begin{picture}(150,150)
\put(-20,0){
\psfig{figure=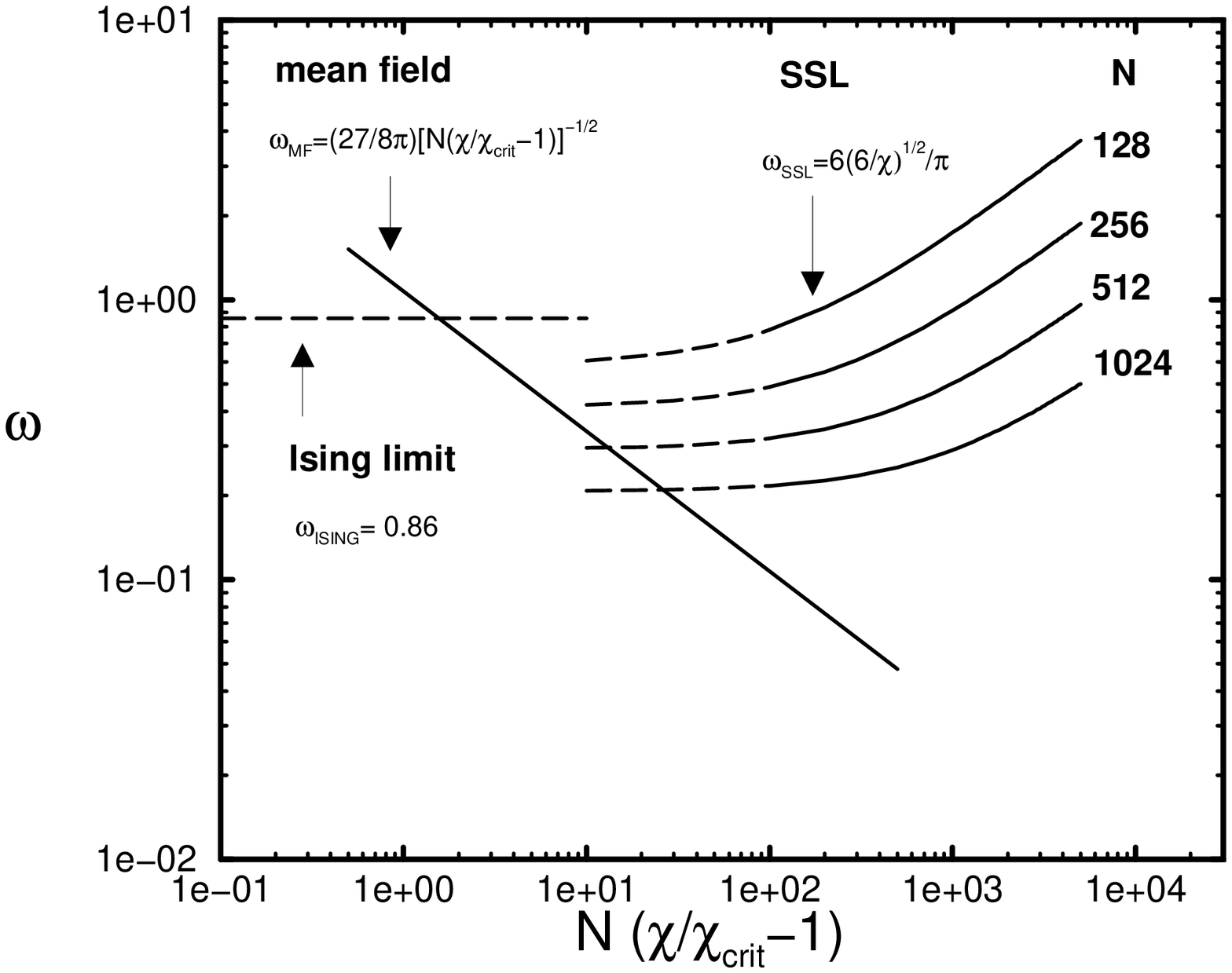,width=160mm,height=140mm}
}

\end{picture}
\vfill
\normalsize
{\tt
\noindent
Figure 2\\
Werner et al\\ JCP
}

\newpage
\pagestyle{empty}

\LARGE
\unitlength=1mm
\begin{picture}(150,150)
\put(-20,0){
\psfig{figure=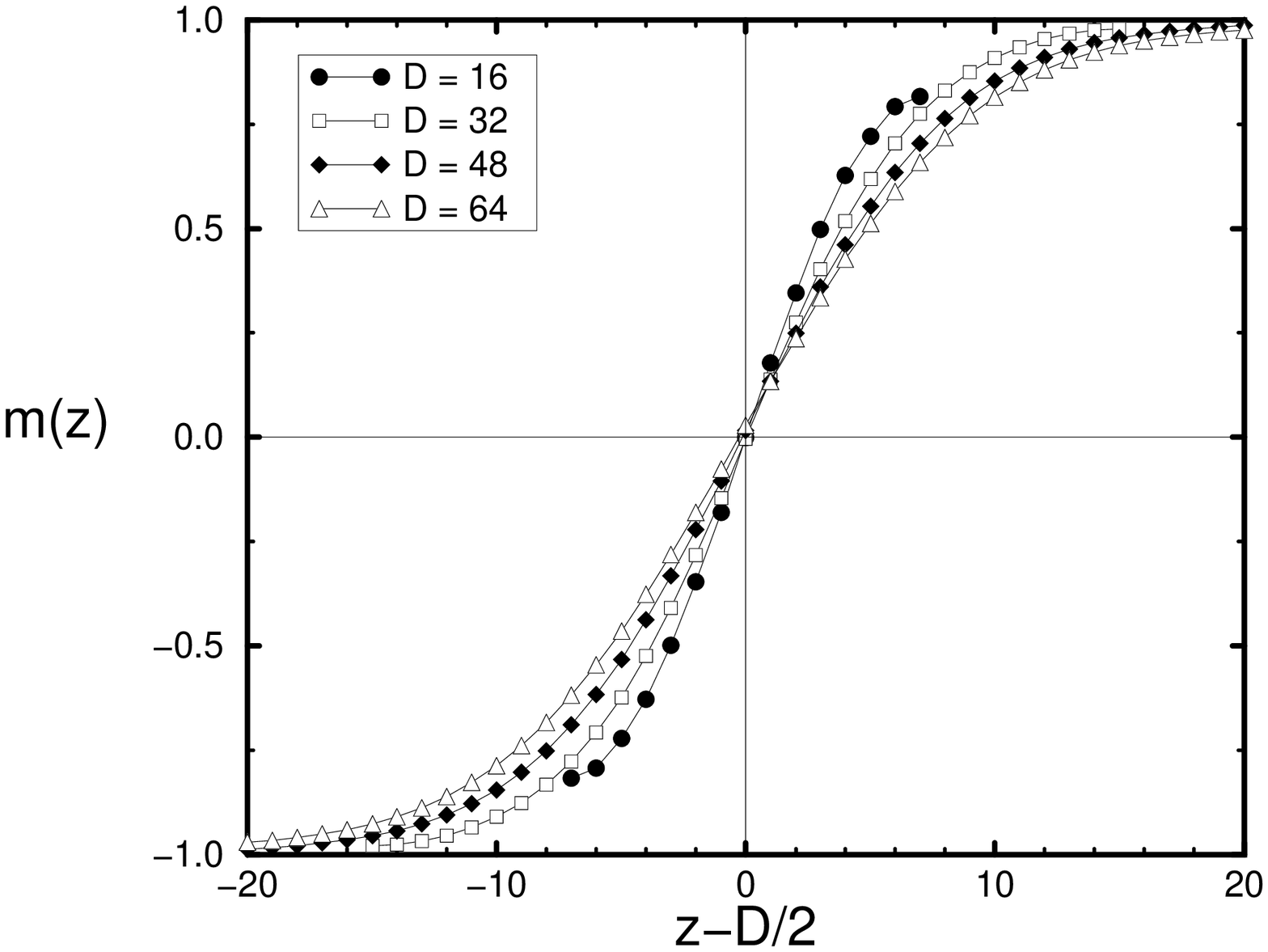,width=160mm,height=140mm}
}
\end{picture}
\vfill
\normalsize
{\tt
\noindent
Figure 3\\
Werner et al\\ JCP
}

\newpage
\pagestyle{empty}

\LARGE
\unitlength=1mm
\begin{picture}(150,150)
\put(-20,0){
\psfig{figure=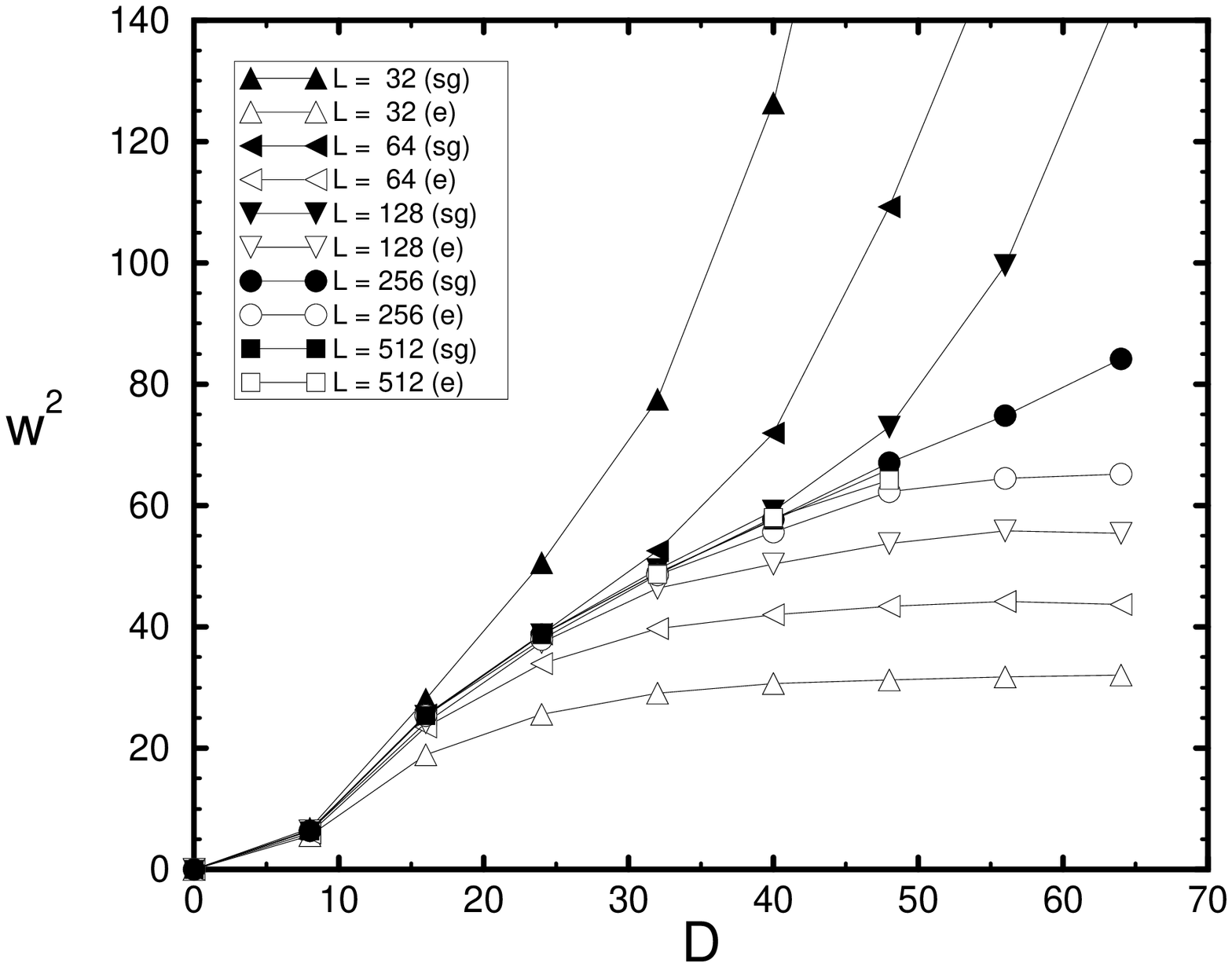,width=160mm,height=140mm}
}
\end{picture}
\vfill
\normalsize
{\tt
\noindent
Figure 4\\
Werner et al\\ JCP
}

\newpage
\pagestyle{empty}

\LARGE
\unitlength=1mm
\begin{picture}(150,150)
\put(-20,0){
\psfig{figure=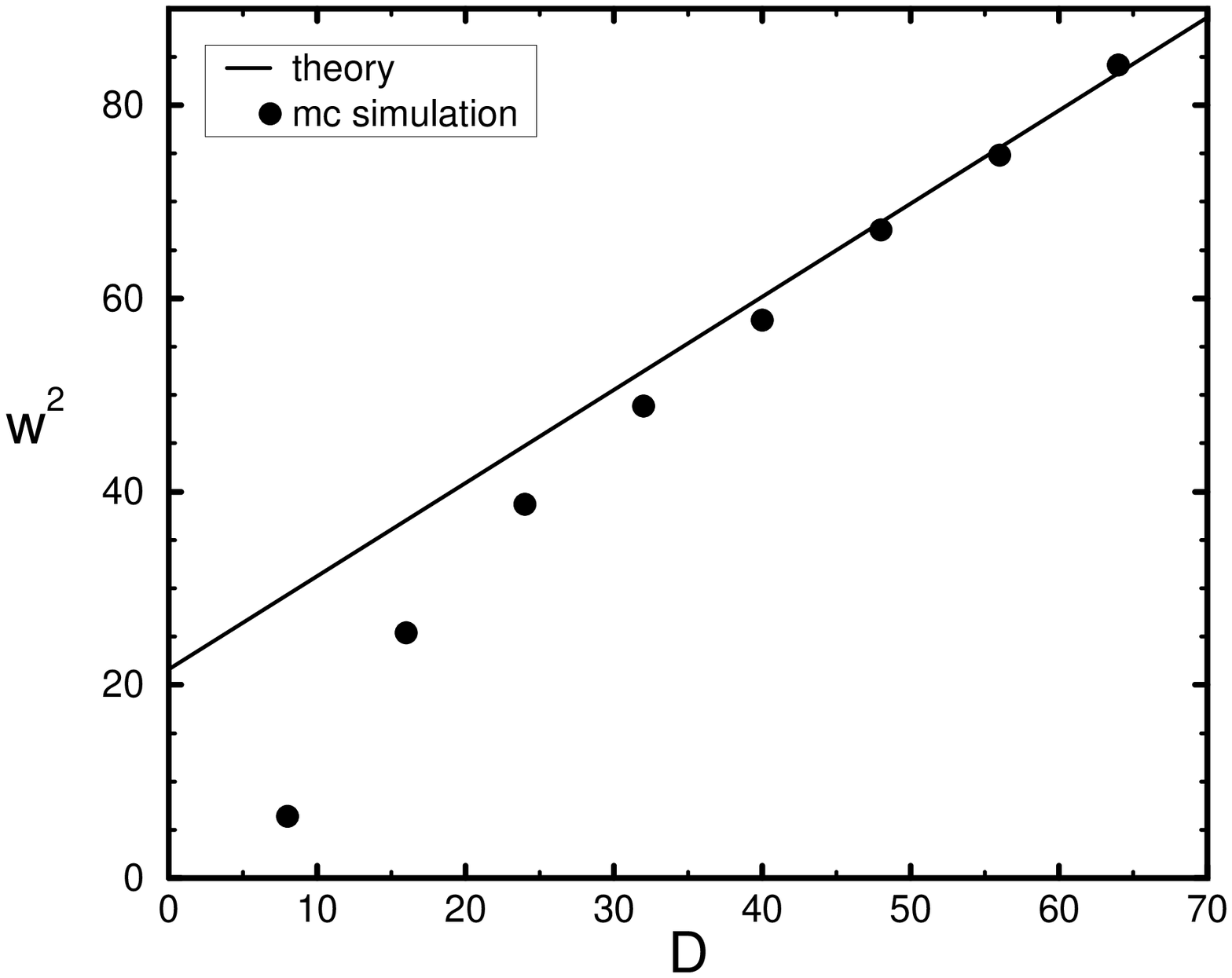,width=160mm,height=140mm}
}
\end{picture}
\vfill
\normalsize
{\tt
\noindent
Figure 5a\\
Werner et al\\ JCP
}

\newpage
\pagestyle{empty}

\LARGE
\unitlength=1mm
\begin{picture}(150,150)
\put(-20,0){
\psfig{figure=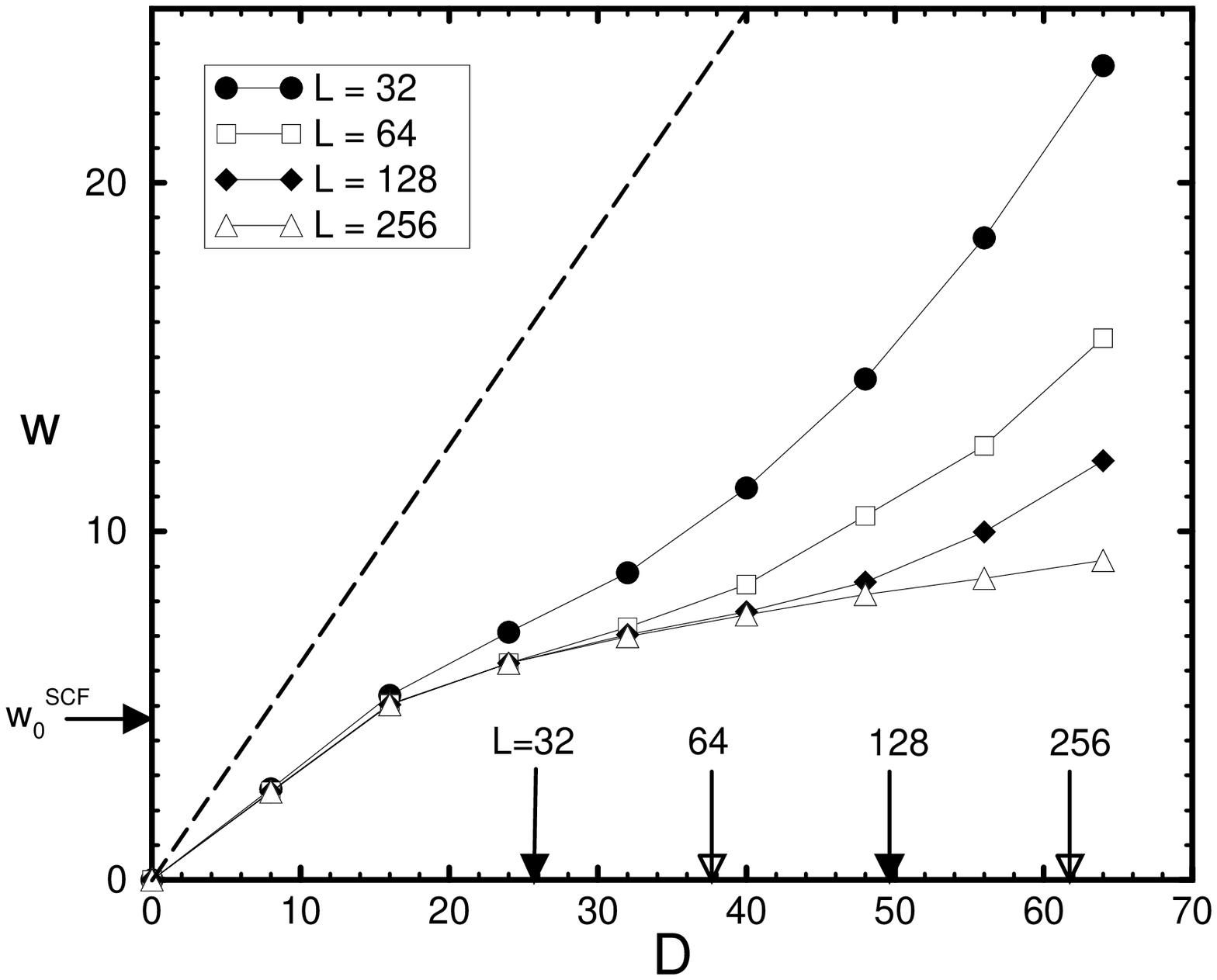,width=160mm,height=140mm}
}
\end{picture}
\vfill
\normalsize
{\tt
\noindent
Figure 5b\\
Werner et al\\ JCP
}

\newpage
\pagestyle{empty}

\LARGE
\unitlength=1mm
\begin{picture}(150,150)
\put(-20,0){
\psfig{figure=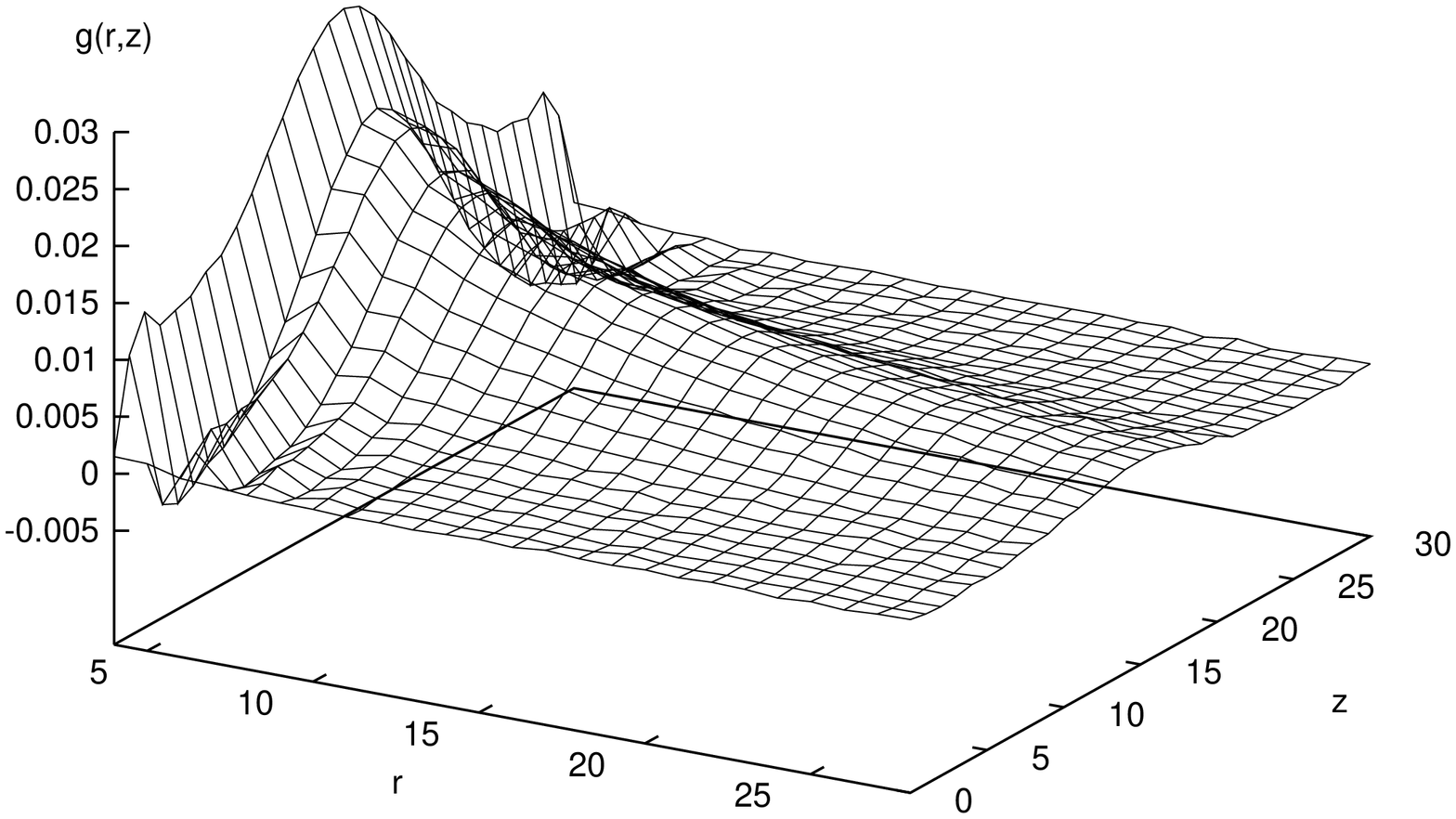,width=160mm,height=140mm,angle=270}
}
\end{picture}
\vfill
\normalsize
{\tt
\noindent
Figure 6\\
Werner et al\\ JCP
}

\newpage
\pagestyle{empty}

\LARGE
\unitlength=1mm
\begin{picture}(150,150)
\put(-20,0){
\psfig{figure=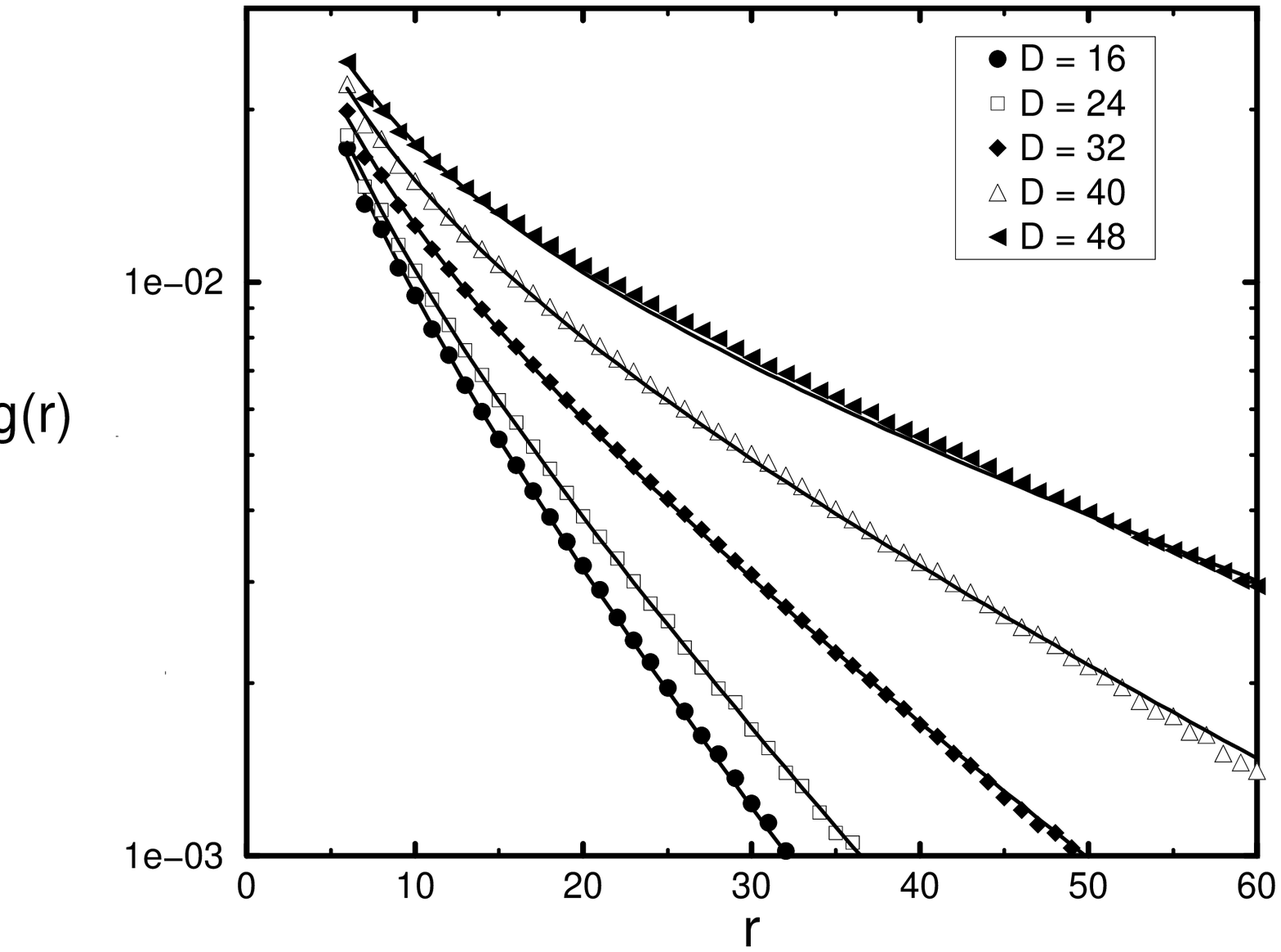,width=160mm,height=140mm}
}
\end{picture}
\vfill
\normalsize
{\tt
\noindent
Figure 7\\
Werner et al\\ JCP
}

\newpage
\pagestyle{empty}

\LARGE
\unitlength=1mm
\begin{picture}(150,150)
\put(-20,0){
\psfig{figure=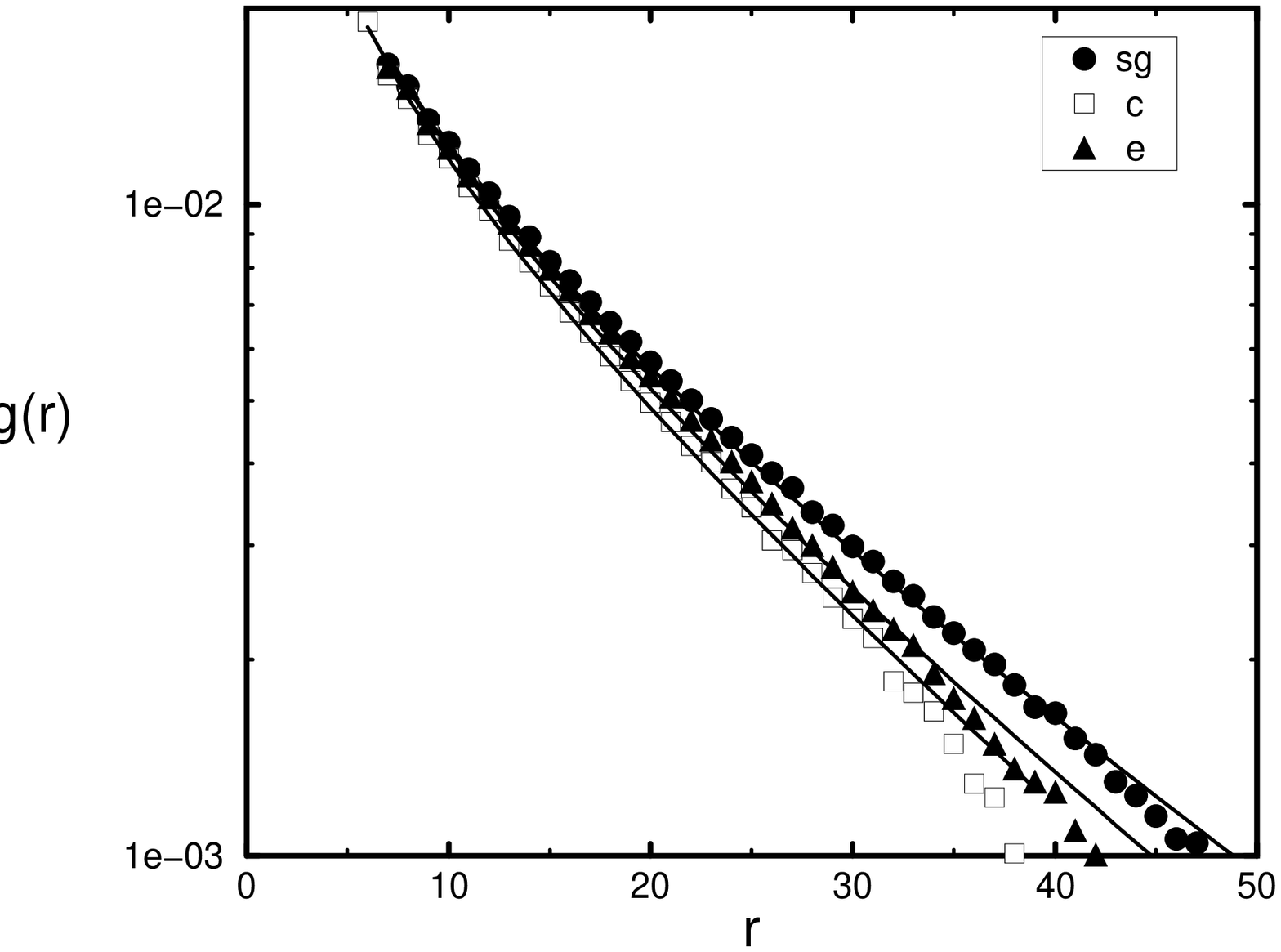,width=160mm,height=140mm}
}
\end{picture}
\vfill
\normalsize
{\tt
\noindent
Figure 8\\
Werner et al\\ JCP
}

\newpage
\pagestyle{empty}

\LARGE
\unitlength=1mm
\begin{picture}(150,150)
\put(-20,0){
\psfig{figure=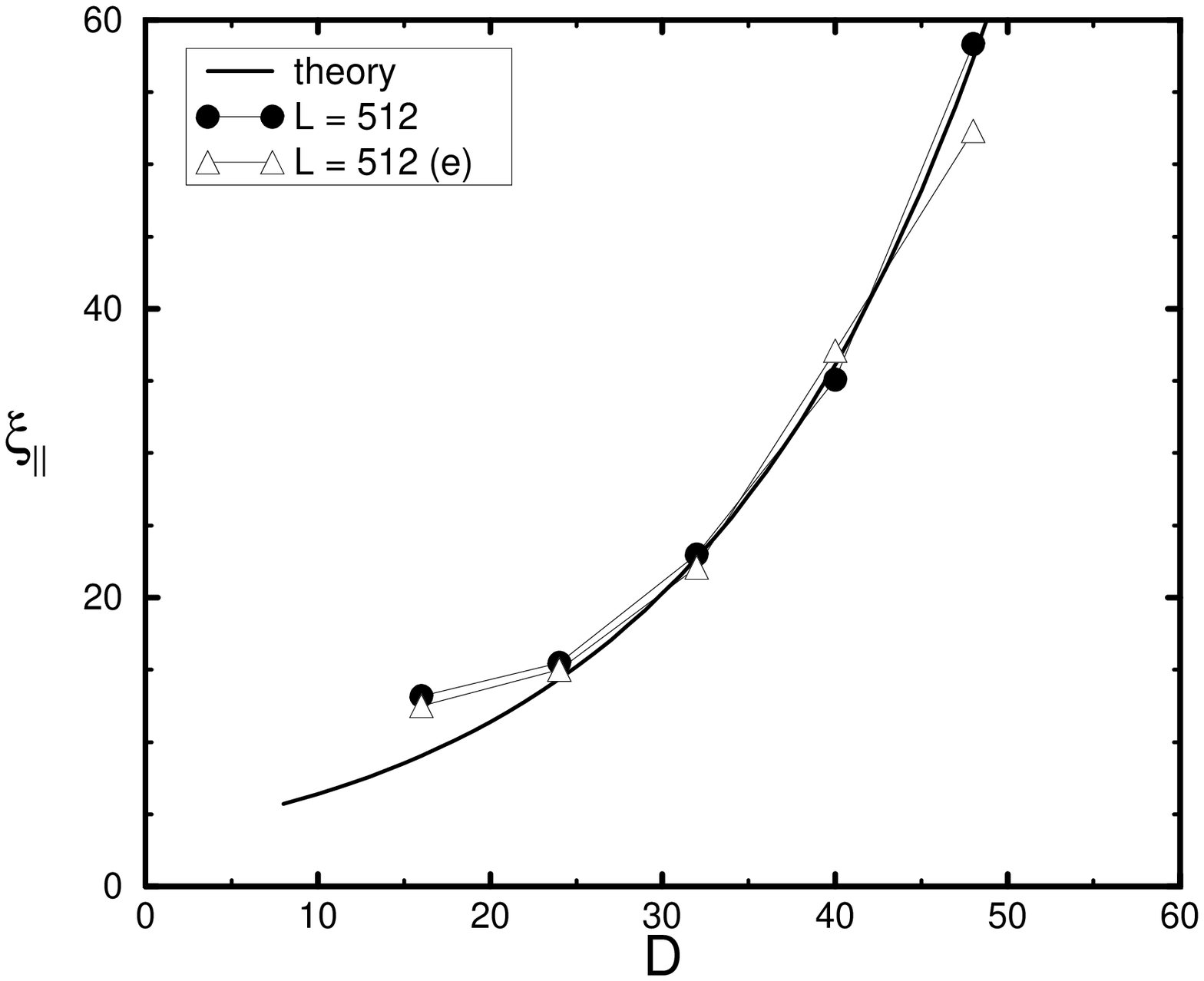,width=160mm,height=140mm}
}
\end{picture}
\vfill
\normalsize
{\tt
\noindent
Figure 9\\
Werner et al\\ JCP
}

\newpage
\pagestyle{empty}

\LARGE
\unitlength=1mm
\begin{picture}(150,150)
\put(-20,0){
\psfig{figure=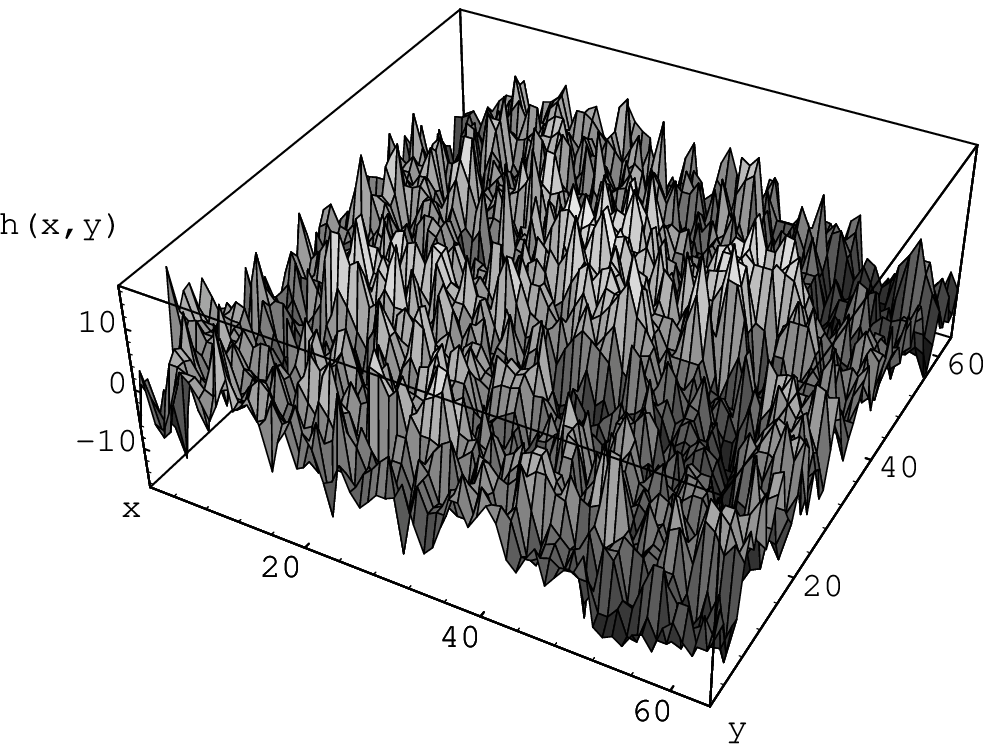,width=160mm,height=140mm}
}
\end{picture}
\vfill
\normalsize
{\tt
\noindent
Figure 10a\\
Werner et al\\ JCP
}

\newpage
\pagestyle{empty}

\LARGE
\unitlength=1mm
\begin{picture}(150,150)
\put(-20,0){
\psfig{figure=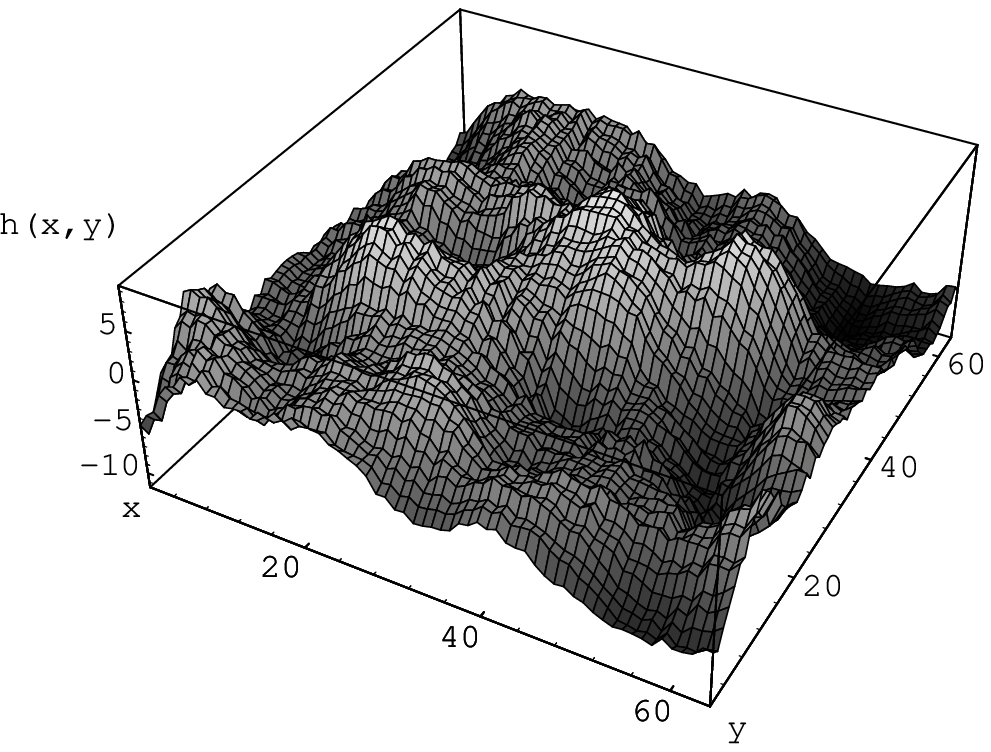,width=160mm,height=140mm}
}
\end{picture}
\vfill
\normalsize
{\tt
\noindent
Figure 10b\\
Werner et al\\ JCP
}

\newpage
\pagestyle{empty}

\LARGE
\unitlength=1mm
\begin{picture}(150,150)
\put(-20,0){
\psfig{figure=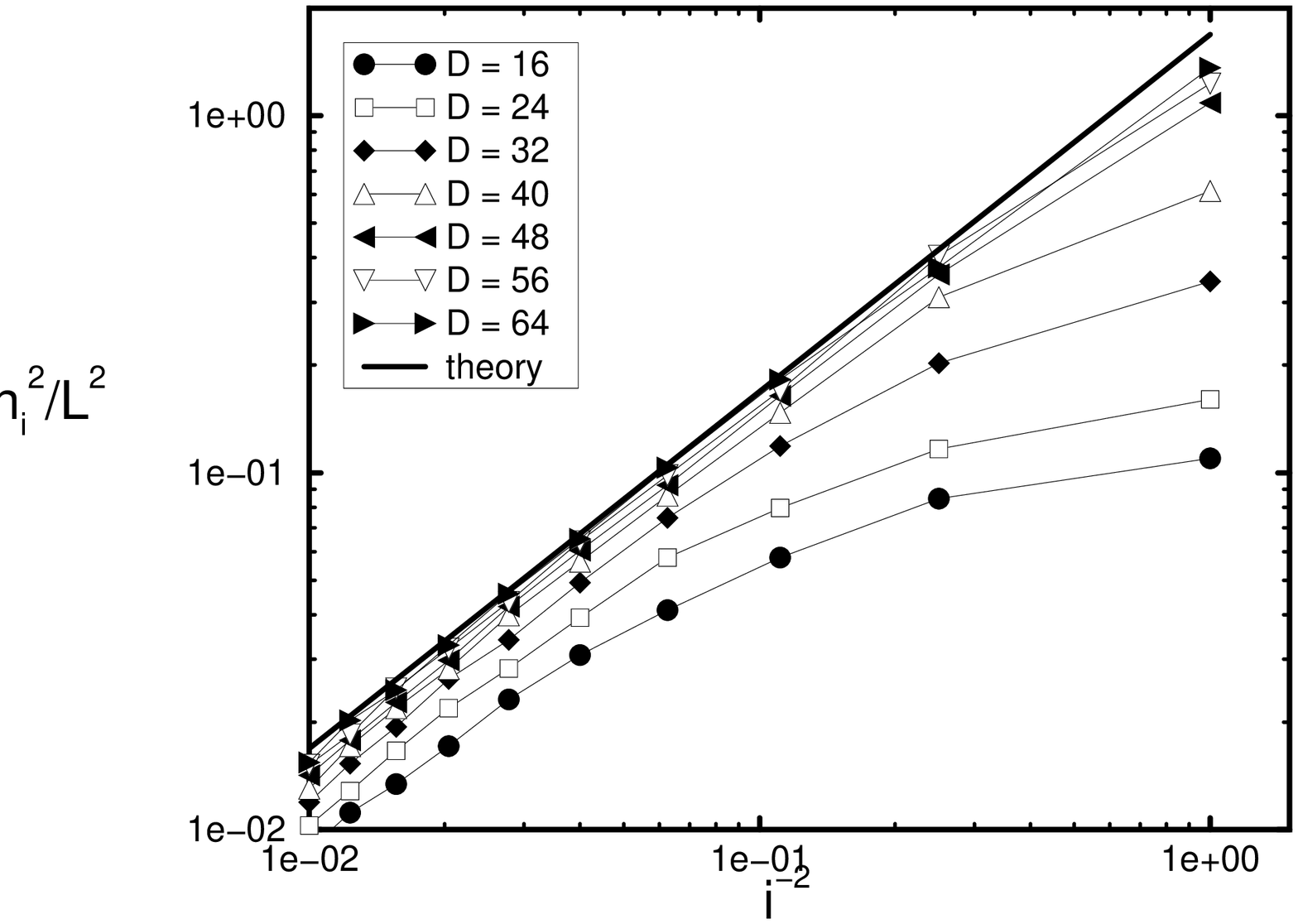,width=160mm,height=140mm}
}
\end{picture}
\vfill
\normalsize
{\tt
\noindent
Figure 11\\
Werner et al\\ JCP
}

\newpage
\pagestyle{empty}

\LARGE
\unitlength=1mm
\begin{picture}(150,150)
\put(-20,0){
\psfig{figure=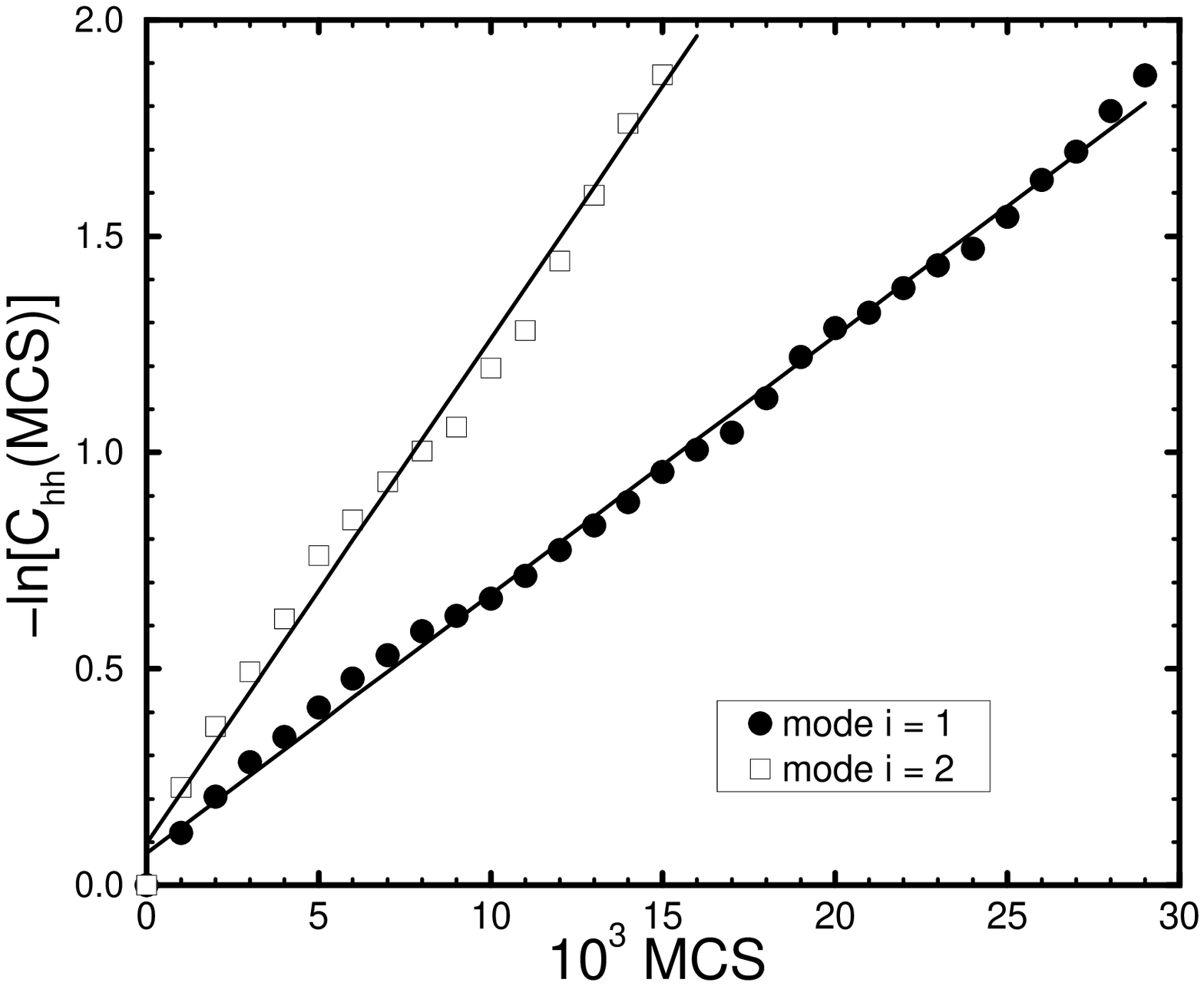,width=160mm,height=140mm}
}
\end{picture}
\vfill
\normalsize
{\tt
\noindent
Figure 12\\
Werner et al\\ JCP
}

\newpage
\pagestyle{empty}

\LARGE
\unitlength=1mm
\begin{picture}(150,150)
\put(-20,0){
\psfig{figure=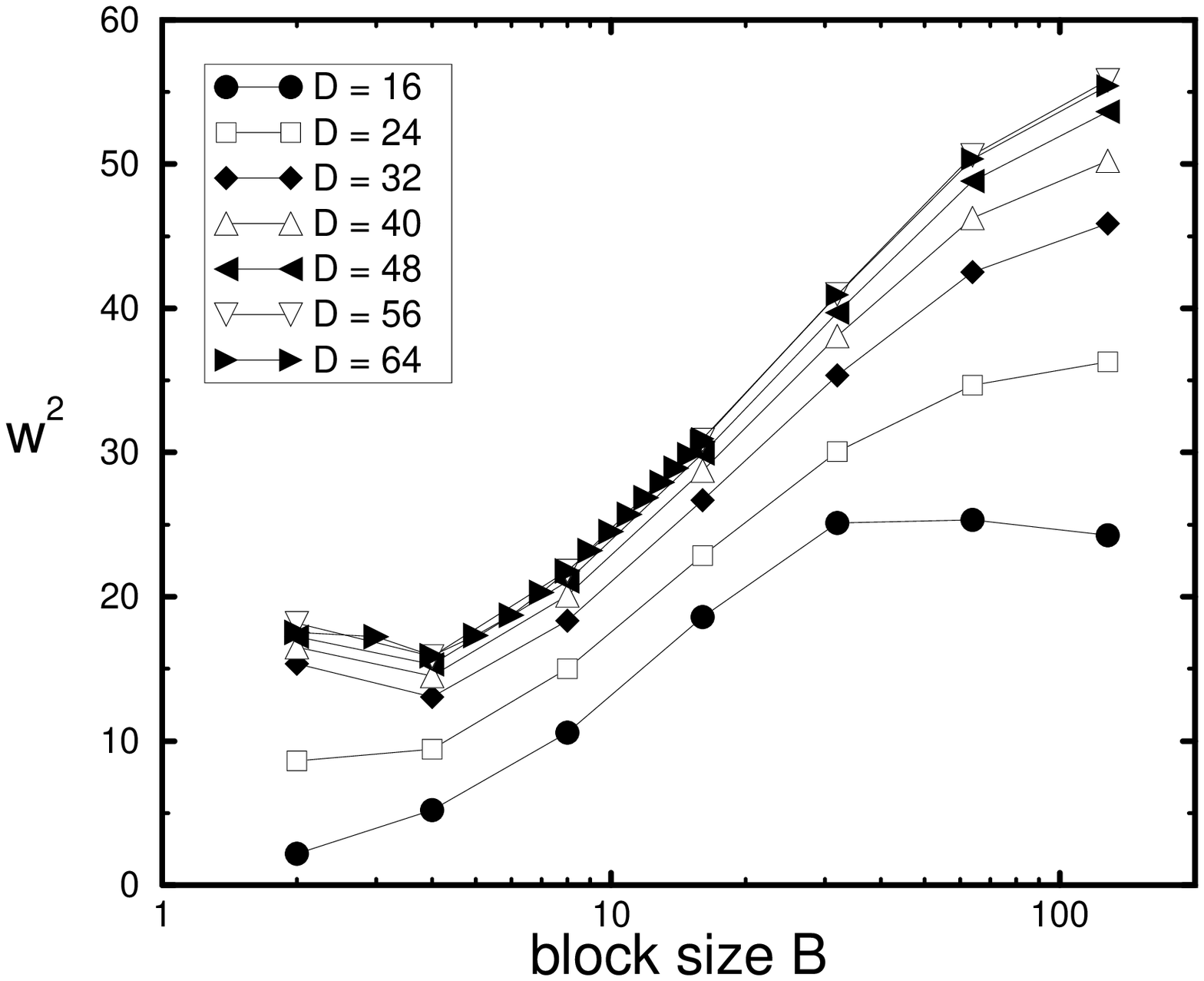,width=160mm,height=140mm}
}
\end{picture}
\vfill
\normalsize
{\tt
\noindent
Figure 13\\
Werner et al\\ JCP
}

\newpage
\pagestyle{empty}

\LARGE
\unitlength=1mm
\begin{picture}(150,150)
\put(-20,0){
\psfig{figure=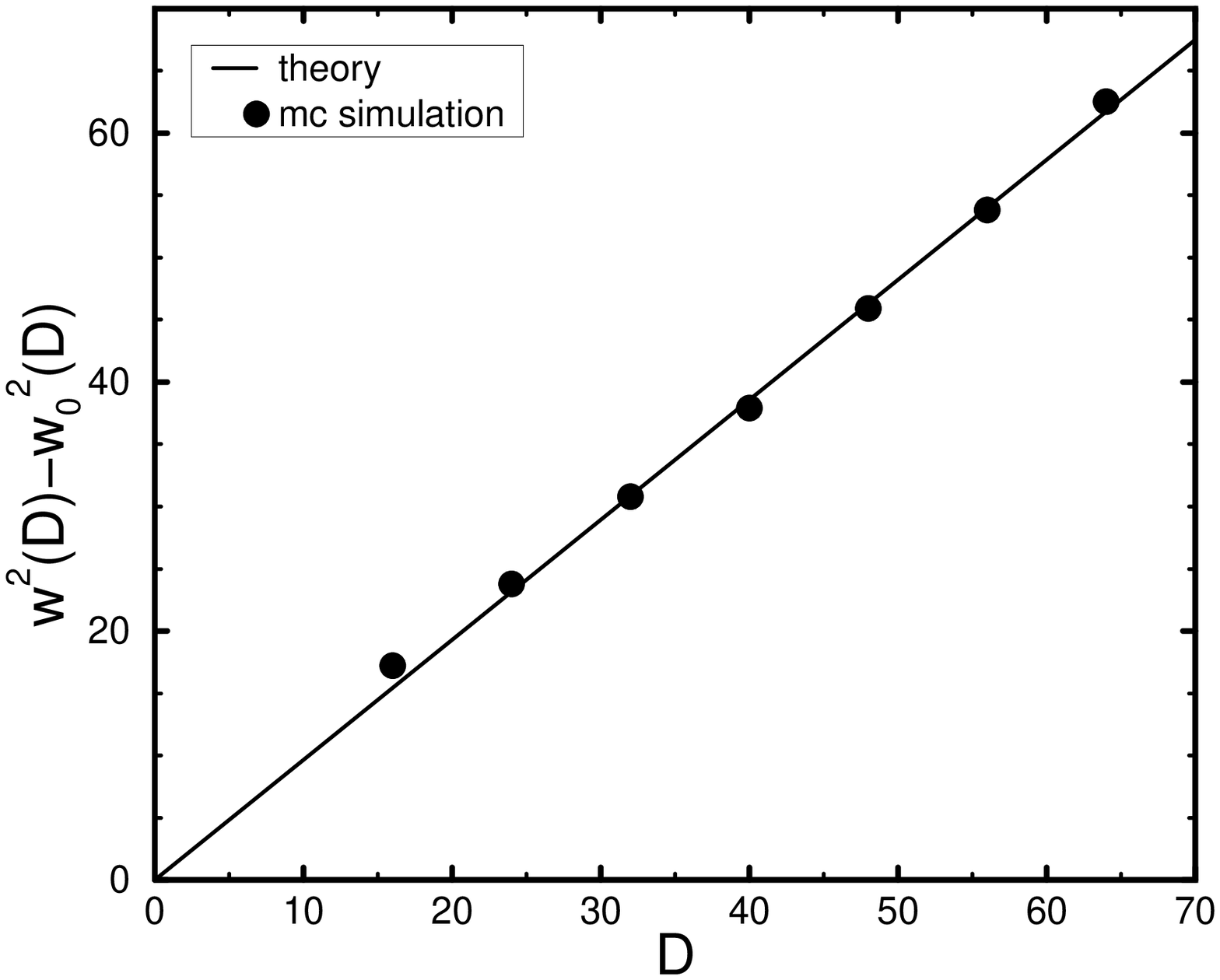,width=160mm,height=140mm}
}
\end{picture}
\vfill
\normalsize
{\tt
\noindent
Figure 14\\
Werner et al\\ JCP
}

\newpage
\pagestyle{empty}

\LARGE
\unitlength=1mm
\begin{picture}(150,150)
\put(-20,0){
\psfig{figure=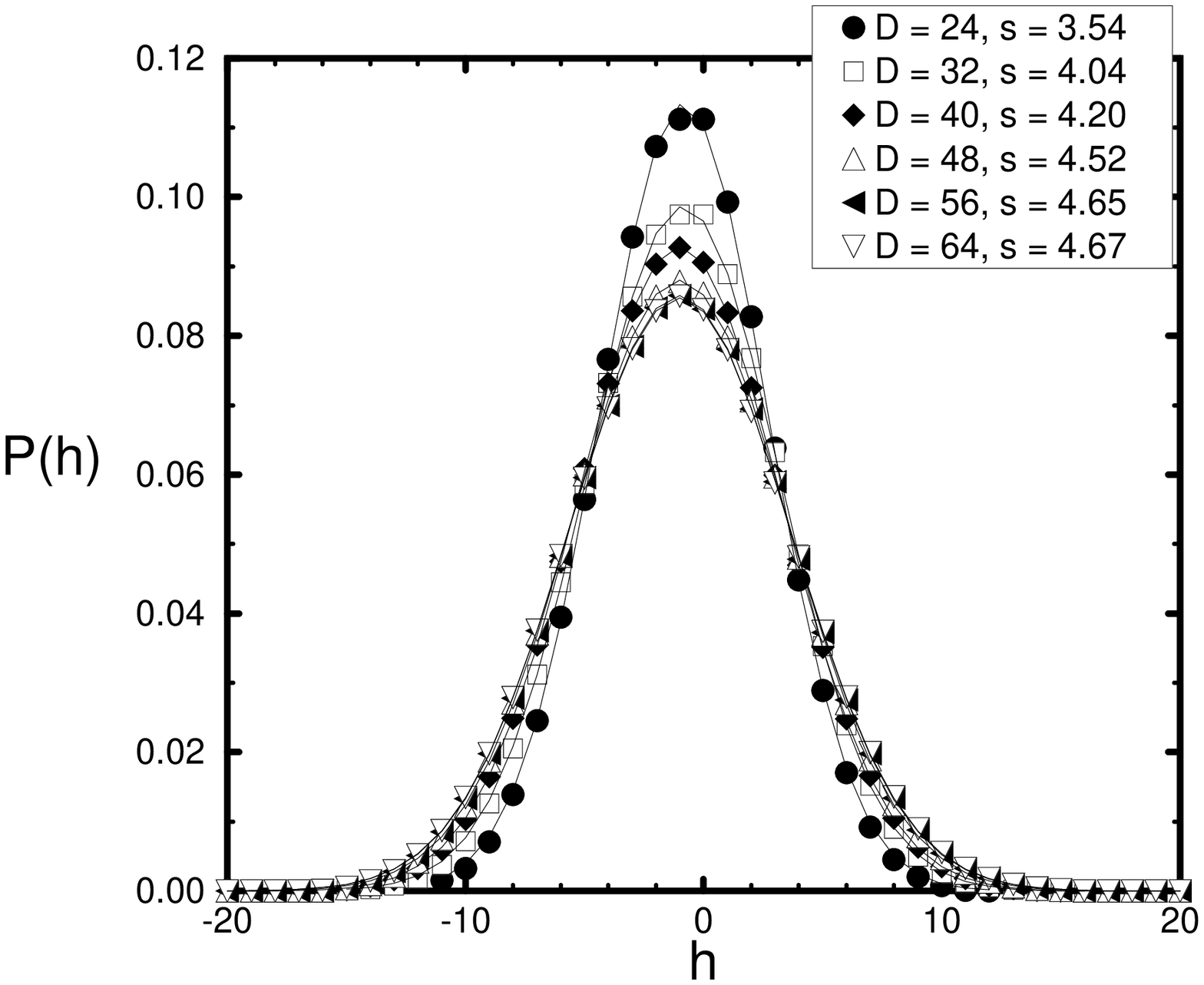,width=160mm,height=140mm}
}	
\end{picture}
\vfill	
\normalsize	
{\tt	
\noindent
Figure 15\\	
Werner et al\\ JCP	
}

%% file: main.bbl
\begin{thebibliography}{99}
\bibitem{ftnt}
In the semi-grand-canonical ensemble, the fluctuations of the
average interfacial position $\overline{h}$ in a configuration are 
according to Eq.28 given by
\[
\langle \overline{h}^2 \rangle = \frac{\xi_{||}^2}{L^2 \sigma}.
\]
For $L \to \infty$, this contribution vanishes, and both
canonical and semi-grand-canonical ensemble yield the same
interfacial width.
For finite $L$, however, the fluctuations of the average position increase
the width.
In the limit $L \ll \xi_{||}$, $\langle \overline{h}^2 \rangle$ is of
the order of $D^2$ itself, and thus $w = D/2$ for $L / \xi_{||} \to 0$.
\bibitem{39}
We call a mixture symmetric if its Hamiltonian (and hence its free energy)
in the bulk is invariant against an interchange of A and B. This
is the case e.g. for the Ising spin model of binary mixtures (for zero
``magnetic field'' the Hamiltonian $\sum_{\langle i,j \rangle} J_{ij} S_i S_j$
is invariant against a reversal of the signs of all spins $S_i$ which may take
values $S_i = \pm 1$, the two signs corresponding to lattice sites $i$ being
taken by A and B, respectively).
``Antisymmetric walls'' mean in the Ising spin representation that the
wall magnetic fields $H^{(l)} (z)$, $H^{(r)} (z)$ due to the left (right)
walls satisfy the condition $H^{(l)} (z) = -H^{(r)} (D-z)$,
choosing the z-axis normal to the interface, and the origin at the left wall
in Fig.1a.
\end{thebibliography}
